\documentclass[letterpaper,11pt]{report}

\usepackage{fullpage}
\usepackage{verbatim}
\usepackage{cite}
\usepackage{setspace}
\usepackage{fancyhdr}
\usepackage{algorithm}
\usepackage{algorithmic}

\usepackage[small]{caption}
\usepackage{graphics}
\usepackage{color}
\usepackage{subfigure}
\usepackage{slashbox}
\usepackage{hyperref}

\usepackage{epstopdf}
\usepackage{amsmath}
\usepackage{multirow}

\usepackage[dvips]{graphicx}

\usepackage[margin=1in]{geometry}
\usepackage{multirow,tabularx}
\newcolumntype{Y}{>{\centering\arraybackslash}X}

\def\title{Thesis proposal}
\begin{document}

\def\addrone{Your address}
\def\addrtwo{Your city}

\def\degree{M.Tech. in Computer Science with Specialization in Information Security}

\def\submissiondate{April 15, 2014}

\def\supervisorone{Dr. Ponnurangam Kumaraguru}

\def\supervisortwo{Dr. Vinayak Naik}

\def\supervisorthree{Mr. Sachin Gaur}

\thispagestyle{empty}

\begin{center}

{\LARGE \bf {Exploration of gaps in Bitly's spam detection and relevant counter measures}

 }
 \vspace{.3in}

 {\Large{Student Name: Neha Gupta}} \\
 \vspace{.1in}
IIIT-D-MTech-CS-IS-14-MT12009 \\

April 15, 2014 \\

    \vspace{.35in}

  \vspace{.25in}

{Indraprastha Institute of Information Technology\\
New Delhi}

\vspace{.35in}  {\underline{Thesis Committee} \\ \supervisorone                ~(Chair) \\ \supervisortwo\\ \supervisorthree } \\ \vspace{.35in}

 {Submitted in partial fulfillment of the requirements \\for the Degree of M.Tech. in Computer Science, \\ with specialization in Information Security}

\vspace{.2in}

\copyright 2014 IIIT-D-MTech-CS-IS-14-MT12009 \\ All rights reserved \\
\vspace{.8in}

\end{center}

\newpage

\pagestyle{empty}
\vspace*{7.1in}
\newpage

\begin{center}
\section*{Certificate}\label{section:certificate}
\end{center}
This is to certify that the thesis titled \textbf{``Exploration of gaps in Bitly's spam detection and relevant counter measures"} submitted by \textbf{Neha Gupta} for the partial fulfillment of the requirements for the degree of \emph{Master of Technology} in \emph{Computer Science \& Engineering} is a record of the bonafide work carried out by her under my guidance and supervision in the Security and Privacy group at Indraprastha Institute of Information Technology, Delhi. This work has not been submitted anywhere else for the reward of any other degree. \\ \vspace{0.5in}

\textbf{Advisor Name}\\Dr. Ponnurangam Kumaraguru

\textbf{Indraprastha Institute of Information Technology, New Delhi}

\begin{abstract}
Existence of spam URLs over emails and Online Social Media (OSM) has become a growing phenomenon. To counter the dissemination issues associated with long complex URLs in emails and character limit imposed on various OSM (like Twitter), the concept of URL shortening gained a lot of traction. URL shorteners take as input a long URL and give a short URL with the same landing page in return. With its immense popularity over time, it has become a prime target for the attackers giving them an advantage to conceal malicious content. Bitly, a leading service in this domain alone shortens close to 80 million links each day, and marks 2-3 million as suspicious every week.~\footnote{\url{http://blog.bitly.com/post/138381844/spam-and-malware-protection}} Some recent research highlights that services from Bitly are being exploited heavily to carry out phishing attacks, work from home scams, pornographic content propagation, etc. In year 2012, one major attack happened in which the U.S. federal government's official short link service \emph{usa.gov} (in collaboration with Bitly) was hijacked to spread work from home scam.~\footnote{\url{http://www.pcworld.com/article/2012800/spammers-abuse-gov-url-shortener-service-in-workathome-scams.html}} Such attacks which targets seemingly secure and highly trusted web sources look alarming and also bring to light the massive impact of exploiting the shortening services. All this imposes additional performance pressure on Bitly and other URL shorteners to be able to detect and take a timely action against the illegitimate content. It therefore becomes important to inspect and identify the root cause and gaps in the implementation of Bitly leading to such attacks. Over the years, multiple defense mechanisms have been set up to handle traditional long URL spam but detection of short URL spam at zero hour still remains a challenging task. 

In this study, we analyzed a dataset marked as suspicious by Bitly in the month of October 2013 to highlight some ground issues in their spam detection mechanism. Our results reveal the inefficiency of Bitly in using some spam detection services that it claims to use. We also show as to how a suspicious Bitly account goes unnoticed despite of a prolonged recurrent illegitimate activity. Bitly only displays a warning page on identification of suspicious links, but we observed this approach to be weak in controlling the overall problem of spam. In addition, we identified some short URL based features and coupled them with two domain specific features to classify a Bitly URL as malicious / benign. Short URL based feature set that we used comprises of click dependent as well as click independent metrics, thus our algorithm can also identify a malicious Bitly URL even before it is actually clicked. The proposed solution is independent of any available blacklists or lexical URL based features. We used standard machine learning classification techniques and were able to detect malicious Bitly URLs with a maximum accuracy of 86.41\%. Although our algorithm is designed specific to Bitly, but we believe that it can be easily extended and used by any other URL shortening services. To the best of our knowledge, this is the first attempt to underline loopholes in security mechanisms of the most popular URL shortening service by analyzing only content the service itself marks as suspicious, and proposing a suitable countermeasure.~\footnote{We have also interacted with `Brian Eoff' (Lead Data Scientist at Bitly) by sharing our analysis and getting his reactions.}
\end{abstract}
\newpage
\pagestyle{empty}
\newpage

\section*{Acknowledgments}\label{section:acknowledgments}
\pagestyle{plain}
\pagenumbering{roman}
First and foremost I offer my deepest gratitude to my advisor, Dr. PK, who has supported me throughout my thesis. His patience and knowledge provided me the room to work in my own way. I attribute the level of my Masters degree to his encouragement and effort without which this Thesis would not have been completed or written. One simply could not wish for a better or friendlier advisor. Besides my advisor, I would like to thank my esteemed committee members Dr. Vinayak Naik and Mr. Sachin Gaur for agreeing to evaluate my thesis. My sincere thanks also goes to Anupama Aggarwal for shepherding me and spending her valuable time to review my thesis. I also thank Bitly and particularly Brian David Eoff (senior data scientist) and Mark Josephson (CEO) for sharing the data with us.

I would also like to express my sincere gratitude to CERC at IIIT-Delhi for providing me an exposure to share my ideas with experts from different parts of the world. I thank my fellow lab-mates from Precog research group at IIIT- Delhi for all their encouragement and insightful comments. Last but not the least, I would like to thank all my family members and friends who encouraged and kept me motivated throughout the project.
\\
\\

\newpage

\tableofcontents
\listoffigures
\listoftables

\newpage

\newpage

\newpage
\mbox{}


\chapter{Research Motivation and Aim}\label{chapter:Research Motivation and Aim}
\setcounter{page}{1}
\pagenumbering{arabic}
\onehalfspacing
\section{Research Motivation}
\subsection{URL shortening services: Background and Use}
URL shortening is a technique of mapping a long Uniform Resource Locator (URL) to a short URL redirecting to the same page. Initially, the concept was used to prevent breaking of complex URLs while copying text, to accommodate long URLs without line breaks, and for smooth dissemination of content. Usage of this service nowadays has become a trend in Online Social Media (OSM) like Facebook and Twitter. Content length restriction imposed by various OSMs (e.g. Twitter's 140 character limit) helped popularize their use even further. In order to accommodate more content in their tweet or Facebook posts, users prefer to compress their long URL using the existing services and embed this URL in their text. Some top URL shortening services like \textit{bit.ly}~\footnote{\url{https://bitly.com/}} and \textit{goo.gl}~\footnote{\url{https://goo.gl/}} also track the URLs and provide real time click traffic analysis. Although these services are created to comfort the users, spammers found their ways to target and misuse the facility for their own good.
\subsection{URL shortening services: Abuse}
URL shorteners does not only reduce the content length but also help obfuscate the actual URL behind a shortened link. Spammers take advantage of this obfuscation to misguide the audience and make quick money by posting malicious links on OSM. These malicious links can be: i) \textit{spam} - irrelevant messages sent to large number of people online, ii) \textit{scam} - online fraud to mislead people, iii) \textit{phishing} - online fraud to get personal user credentials, or iv) \textit{malware} - auto downloadable content to damage the system. According to a threat activity report by Symantec in year 2010 \cite{20}, around 65\% malicious URLs on OSM are shortened URLs. Another study in year 2012 reveals that around 80\% of shortened URLs contained spam-related content \cite{2}. Research by a URL shortener \textit{yi.tl} reveals that because of deep penetration of spam, 614 out of 1,002 URL shortening services became dead in year 2012.~\footnote{\url{http://www.webpronews.com/study-claims-61-of-url-shorteners-are-dead-2012-05}} A recent article in year 2013 also highlights that Facebook spammers makes close to 200 million dollar just by posting these shortened links to lure users \cite{21}. 

Bitly, launched in year 2008 is one of the most popular URL shortening and curating services on the web \cite{18}. It gained major traction when Twitter started to use it as a default URL shortener in year 2009 before the launch of its own service, \textit{t.co} in the year 2011.~\footnote{\url{https://support.twitter.com/articles/109623-about-twitter-s-link-service-http-t-co}} Bitly provides an interface to its users to create an account and shorten the links. Each link shortened by a user has a unique \textit{global hash} (an aggregated identifier corresponding to a link). Such shortened links, known as \textit{bitmarks} can then be saved, tracked, and shared. Users are also allowed to connect any number of Facebook / Twitter accounts with their Bitly accounts. Simple interface and ease of social media connection makes the task of shortening and sharing a link very convenient. With 80 million new links shortened on Bitly each day and 8 billion clicks each month, spammers have also been exploiting the service to a great extent.~\footnote{\url{http://www.enterprise.bitly.com/about-us/}} In early 2013, a news article reported the spread of phishing attacks on Twitter using malicious Bitly links.~\footnote{\url{http://news.softpedia.com/news/Twitter-Phishing-Scam-This-Profile-Is-Spreading-Nasty-Blogs-Around-About-You-318618.shtml}} In this attack, Direct Messages (DM) of Twitter users were targeted with malicious content that read `\textit{FYI this profile on twitter bit.ly/Uf0Cg9 is spreading nasty blogs around about you}'. Large number of users receiving such a message fell in the trap and clicked on the link, which actually redirected to a website that replicated Twitter's login page. Victims were then misled that their session was expired and were made to login again, unknowingly revealing their Twitter credentials to the attacker. Impact of the attack was such that Twitter announced a temporary restriction on sending shortened links including Bitly in DM.~\footnote{\url{http://www.digitaltrends.com/social-media/twitter-may-be-banning-links-in-dms/\#!CvRAI}}

\begin{figure}[h!]
     \begin{center}
	\includegraphics[scale = 0.65]{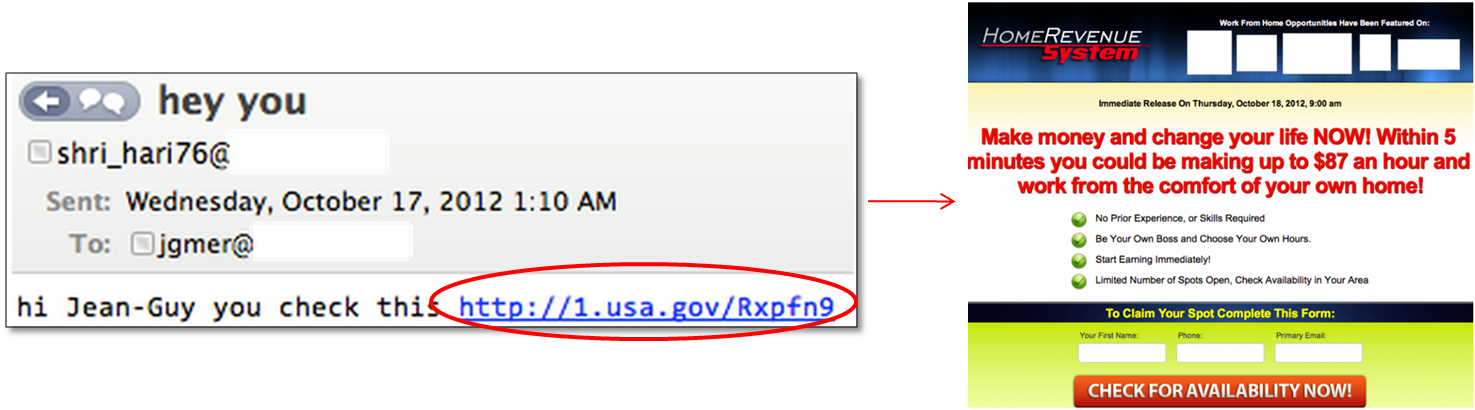}
     \caption{Attack on short URL domain \textit{1.usa.gov} leading to work from home scam.}
	\label{usaGovAttack}
     \end{center}
\end{figure}

Fig~\ref{usaGovAttack} shows another attack where spammers abused the open redirect vulnerabilities of a popular and legitimate domain belonging to the U.S. federal government, in collaboration with Bitly. The hijacked domain \textit{1.usa.gov} which redirected a user to illegitimate work from home scam website received around 43,049 clicks from 124 countries within a week, contributing to 15.1\% of all \textit{1.usa.gov} URLs.~\footnote{\url{http://www.pcworld.com/article/2012800/spammers-abuse-gov-url-shortener-service-in-workathome-scams.html}} This clearly highlights that even branded short URLs are not safe from exploits. To raise awareness about other dangers associated with URL shorteners, a researcher from University of Tulsa developed his own shortening service (\textit{d0z.me}) and conducted a distributed denial of service attack (DDoS) experiment. An attacker could shorten a link using this service and mention the server to be attacked. \textit{d0z.me} in turn gave a malicious short URL, which when clicked introduces a DDoS attack on the targeted server as long as the user remains on the website. This malicious service witnessed significant number of hits showing the impact and extent of vulnerabilities of URL shortening services. In October 2013, Bitly also experienced a massive DDoS attack rendering complete shutdown of its services for close to 7 hours.~\footnote{\url{http://www.geeknewscentral.com/2013/10/21/twitter-banning-bit-ly-other-url-shortners-on-direct-messages-dm/}} Some of the spammers have also started to build their own URL shortening services to double shorten the malicious links, first with a self created short URL service, then again with a legitimate short URL service to evade security checks. The security researchers from Symantec also found that spammers abused the services from Bitly to propagate sexually suggestive content \cite{16}. 
Unlike other URL shortening services like \textit{goo.gl}, Bitly does not provide a captcha to track the non-human accounts. For protection against spam, Bitly claims to use real-time spam detection services like Google safe-browsing and SURBL, and flags 2-3 millions of the shortened links as spam each week.~\footnote{\url{http://blog.bitly.com/post/138381844/spam-and-malware-protection}} Bitly neither deletes a flagged suspicious link nor suspends the associated user, but displays a warning page whenever such a link is clicked. Such a warning page is by-passable and thus does not completely restrict a user to visit a malicious website. Also, non-deletion of malicious content or account can make it persistent over web. Looking at these detection measures adopted by Bitly but continued existence of illegitimate 
content, it becomes important to understand the security issues associated with the service that deters its efficiency to handle malicious links.

\section{Aim}
In this work, we perform a detailed analysis on a dataset of suspicious Bitly links (directly obtained from Bitly) and their associated attributes to:
\begin{itemize}
 \item Underline the characteristics of domains suspicious Bitly URLs are coming from and the top referrers being targeted
 \item Study the impact of malicious Bitly links on OSM
\item Inspect Bitly's current spam detection techniques and expose the associated security issues
\item Identify a set of Bitly specific features to detect spam
\end{itemize}
To the best of our knowledge, this is the first attempt to inspect only data labeled as suspicious by Bitly to underline security loopholes contributing to such a data and proposing a suitable solution.



\chapter{Literature Review}\label{chapter:Literature review}
\onehalfspacing

URL shortening services takes as input a long URL and generates a short URL hash in return. This short URL so generated redirects to the same long URL but looks random and unrelated to the actual link. Imposed character limit has lead to immense popularity of such services in social media landscape. Due to its ubiquitous usage, these services have been hit by adversaries to obfuscate and disseminate bad content. This section presents the work done in understanding the usage pattern, behavior, use, misuse of short URLs along with some studies dealing with the detection of long malicious URLs. 
\section{Malignant Long URL Characterization / Detection}
A large number of studies have been conducted to understand the propagation of spam on OSM, many of which revealed heavy usage of URLs to spread bad content. Benevenuto et al. in their research identified distinctive features to detect spammers on Twitter for a dataset of 54 million users and 1.9 billion links \cite{8}. After doing a manual annotation to label spammer / non-spammer, they defined content and user behavior based features to classify a spammer and achieved an accuracy of 70\%. They also found \textit{fraction of tweets with URLs} and \textit{average number of URLs in a tweet} to be the top feature set in their classification. This clearly highlights the extent to which URL based metrics alone contribute to spam detection. Researchers  also evaluated the effectiveness of popular blacklists in evading spam and observed it to be very inefficient. In case of Twitter, it becomes even more slow because of the URL shortening services used to obfuscate long URLs. Using these services, a spammer can complicate the process of detection by using chains of multiple shortenings \cite{10}. Grier et al. also pointed out that the clickthrough rate for Twitter spam (0.13\%) is higher than email spam. 

Monarch is another system developed by Thomas et al. which classifies a URL submitted to any web service as malicious or benign in real time \cite{6}. This system relies on the features of URL landing page, i.e. the page content, behavior, hosting infrastructure, pop-ups, plugins and other raw attributes which may sometime not be reachable. Although authors claim the accuracy to be close to 91\%, but heavy text based computation may lead to performance issues. For our classification, we don't rely on any such text based features. In year 2012, Aggarwal et al. also proposed a real time phishing detection system for Twitter, called PhishAri \cite{9}. Authors coupled the Twitter and URL based features to classify phishing tweets and achieved an accuracy of 92.52\%. 
In another real time suspicious URL detection technique on Twitter proposed by Lee et al., authors addresses the problem of conditional URL redirects \cite{11}. Many suspicious websites perform conditional redirection where a normal user is redirected to a malware / phishing website, but a crawler is detected and redirected to a normal website. The intention is to fool web investigators. The authors thus focused on creating a system that would not fall prey to conditional redirection. A combined feature set of \textit{correlated URL redirects} and \textit{tweet context information} was used and an accuracy of 86.3\% was attained. 

\section{Short URL Analysis}
With introduction of short URLs in OSM, a comparative look was also necessary to be able to understand the level of acceptance of the service over long URLs. Kandylas et al. performed a relative study of long and short Bitly URLs propagation on Twitter and found that Bitly links received orders of magnitude more clicks than an equal random set of long URLs \cite{5}. They also observed that the lifetime of spam URLs is shorter as compared to legitimate URLs. 
To further comprehend short URL distinctive characteristics, Antoniades et al. performed an analysis on 2 different datasets \cite{3}. They crawled Twitter to extract URLs and found 50\% of them to be Bitly. Another dataset they collected directly by brute force crawling of 2 URL shortening services. 
Their results reveal that the lifetime of 50\% short URLs exceeds 100 days. They also found a space gain of 95\% using short URLs in their dataset, and observed 54\% performance overhead on using these services. 

Other than this generic analysis, Neumann et al. looked at malicious shortened URLs in emails and highlighted the privacy and security implications of using short URLs \cite{1}. They found a lot of private user information traces associated with short URLs and observed a low spam detection rate for 16 shortening services they analyzed. For Bitly, spam detection rate was found to be 57\% in their experiment, which can not be stated as optimal for the most popular service in this domain. Chhabra et al. also gave an overview of evolving phishing attacks through short URLs on Twitter and found that phishers use URL shorteners not only to gain space but hide their malicious links \cite{13}. Their results shows that most of the tweets containing phishing URLs comes from inorganic (automated) accounts. They also observed a change of focus of phishers from e-commerce websites to OSM and to countries that are developing fast. Later in year 2012, Klien et al. presented the global usage pattern of short URLs by setting up their own URL shortening service over a span of 20 months \cite{2}. They identified few features from their usage log and found short URL usage trends to be different across different countries. Their analysis also revealed 80\% short URL content to be spam related, which can probably be due to the fact that they did not perform their analysis on some top existing and relatively secure URL shortening services. 

In year 2013, Maggi et al. performed a large scale study on 25 million short URLs belonging to 622 distinct URL shortening services \cite{4}. Their results highlights that the countermeasures adopted by these services to detect spam is not very effective and can be easily bypassed. Experimental results from their data shows that most of the shortening services that accept legitimate content initially do not keep a track if content changes to malignant late in time. They also observed that Bitly allows users to shorten malicious links and does not include any initial level lightweight check to prevent bad shortenings (though detects it after some time). Another scheme was proposed by Yoon et al. in year 2013 about using relative words of target URLs in short URLs \cite{14}. This can give hints to user to guess the target URL, making it comparatively safe from phishing attacks.

\section{Malignant Short URL Characterization / Detection}
The only work that presents some short URL based features to detect malicious accounts is given by Wang et al. in year 2013 \cite{12}. In their experiment, they investigated the creators of 600,000 short Bitly URLs and associated click traffic from different countries and referrers. Based on the analysis, they classify a link as spam / non-spam using only 3 click traffic related features with maximum accuracy of 90.81\%. During the classification, they ignored all the short URLs that did not receive any clicks. In addition, their results reveal that legitimate Bitly users also generate spam and most clicks on short malicious URLs comes from popular websites abused by spammers. 

After reviewing all the above techniques, it was evident that a lot of work has been done in the identification and analysis of malicious URLs on the web. Surprisingly, almost no (very little) work has been done in analyzing only the suspicious shortened URLs to expose the gaps in security mechanisms and policies adopted by a specific URL shortening service. Our work significantly differs from the prior studies as we focus on understanding the ground security issues associated with a URL shortener that deters its efficiency to handle bad content.
Though our classification problem looks a little similar to the one proposed by Wang et al., but we look at a more extensive feature set to label a link as malicious / benign. Since we do not only rely on click traffic based features, our algorithm also works for Bitly URLs that do not receive any clicks. This is important since it allows early detection of short malicious URLs much before they target their audience. To the best of our knowledge, the additional short URL feature set that we identified have not been highlighted in the research community till now.




\chapter{Research contribution}\label{chapter:Research contribution}
\onehalfspacing
In the first part of our research, we analyze the general characteristics of malicious Bitly URLs and their impact on OSM. The major findings from this study are:
\begin{itemize}
 \item Malicious users register domains with a definite purpose of spamming using Bitly. Such domains are usually short lived and eventually die out after receiving a significant number of hits. 
 \item There exists large communities propagating bad content through Bitly on Twitter. Such communities grow in size because Bitly imposes no limit on the number of connected OSM accounts. 
\end{itemize}
The next part of our study highlights the weaknesses in Bitly's security policies. Three major loopholes that we identify are - 
\begin{itemize}
 \item Bity is unable to detect malicious links already tracked by popular blacklist services and is not even using the claimed spam detection techniques effectively.
 \item Malicious users exploit Bitly's no account suspension policy and keep shortening only bad URLs. 
 \item A Bitly warning page is not a good enough measure to restrict the overall problem of spam. 
 \end{itemize}
After identifying these security issues with Bitly, we propose a technique to classify a Bitly URL as malicious or benign. The major contributions of this work are -
\begin{itemize}
\item Unlike most of the previous studies, our detection mechanism relies on the combined features of Bitly and some corresponding long URL based features.
\item Our technique can work efficiently irrespective of the number of clicks received by a Bitly URL. 
\item The classification approach works the best to track malicious Bitly links at zero hour, i.e. when no click is received. 
 \end{itemize}
To the best our knowledge, this is the first large scale study to highlight the issues with Bitly's spam detection policies and proposing a suitable solution.

\chapter{Solution Approach}\label{chapter:Solution Approach}
\onehalfspacing
In this section, we discuss the two step methodology used in our study. 
\section{Suspicious short URL characterization and exploration of Bitly's policies in handling spam}
URL shorteners have emerged and gained a special position in link dissemination over web. Bitly is one of the leading services in this domain, so much so that it alone accounts for about half of all URL shorteners on Twitter \cite{3}. Its extensive use and popularity has attracted a lot of spammers to propagate spam. Whenever Bitly identifies a malicious URL, it displays a warning page to its users. To analyze the basic characteristics of short URL spam, we requested Bitly to share with us the links that it marks as suspicious. In response, we received a dataset of 763,160 suspicious Bitly URLs in the month of October 2013. This dataset comprise of the global hash (an aggregated identifier corresponding to a link), associated long URL, and number of warning pages displayed for the global hash. We call this our \textit{link-dataset}. Bitly also provides with a public API~\footnote{\url{http://dev.bitly.com/api.html}} to extract the link and user metrics for a particular short URL. Using the link-dataset as our seed input and publicly available Bitly API, we collected the analytics for 144,851 (18.98\%) links between January 2014 to March 2014 (our data collection is still on). Table~\ref{BitlyAnalytics} presents these analytics. We call this our \textit{link-metric-dataset}. Using this dataset, we analyzed the creation / propagation trend as well as the connected social network impact of the malicious links on Bitly. We also highlighted some major loopholes in the security system and policies used by Bitly.

\begin{table}[h]
\small
 \begin{center}
    \begin{tabular}{|l|l|}
    \hline
    {\bf Short URL metric}     & {\bf Output data}                                                             \\ \hline
    info                 & link creator and creation time                  \\ \hline
    expand               & target long URL                                                         \\ \hline
    clicks               & last 1000 click history                                                 \\ \hline
    referring\_domains   & domains referring click traffic to the given Bitly URL                  \\ \hline
    encoders             & users who encoded the given Bitly URL                                        \\ \hline
    encoder\_info        & Bitly profile name, profile creation date and connected social networks \\ \hline
    encoder\_link\_history & last 100 links shortened by the encoder                                 \\ \hline
    \end{tabular}
    \caption {\label{BitlyAnalytics} Collected Bitly analytics for suspicious links.}
 \end{center}
\end{table}
\section{Malicious Short URL Detection}
After characterizing the suspicious links obtained from Bitly, we observed and identified few discriminating short URL based features for spam. We coupled these features with two domain specific features to label a short URL as malicious or benign using standard machine learning techniques. For training and testing of data, we extracted 34,802 Bitly links from Twitter and used standard blacklists to formulate the ground truth labeled dataset. 

\chapter{Experiments and Analysis}\label{chapter:Experiments and Analysis}
\onehalfspacing
In this section, we focus on the analysis of dataset obtained from Bitly. Our objective in this research is to highlight some characteristics of malicious short URLs and to bring to light the weaknesses in security mechanisms used by Bitly. We attempt to answer some unexplored questions related to short URL spam like: (i) What are the top referrers of suspicious Bitly URLs and characteristics of domains such content is coming from? (ii) Is Bitly using the claimed spam detection services effectively? (iii) Do malicious users take advantage of Bitly's no account suspension policy? (iv) Does a warning page alone help curtail the overall problem of spam? (v) How quick is Bitly in identifying suspicious accounts?
Answering these questions is important to investigate the competence and helpfulness of Bitly in dealing with illegitimate content.

\section{Domain Analysis}
\begin{figure}[h]
     \begin{center}
		\includegraphics[scale = 0.65]{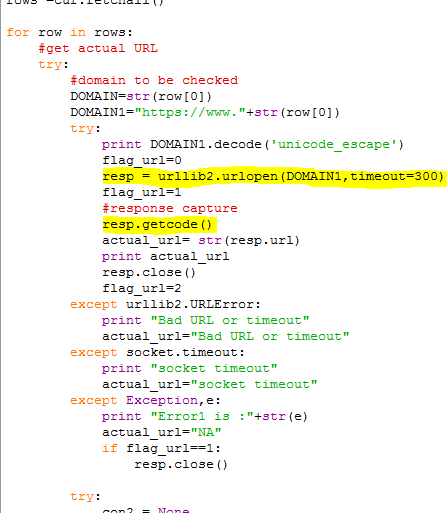}
	\caption{Python code snippet to recheck the presence of each domain after 5 months.}
		\label{domainCheckCode}
     \end{center}
\end{figure}

Quick and easy availability of domains has made the task of a spammer more convenient. Our link-dataset of 763,160 suspicious URLs comprised of 22,038 unique domains. Since our target was to analyze only the malicious domains, we realized that the spammers many-a-times exploit some legitimate popular domains to propagate spam. Keeping this in mind, we used APWG (Anti-Phishing Working Group) whitelist and separated all the legitimate domains from our link-dataset, if any. Here we found 56 exploited domains to be whitelisted. Ignoring these, we created a python crawler (Fig~\ref{domainCheckCode}) and performed a test on the existence of each suspicious domain 5 months after we received our dataset. It was observed that 83.06\% domains no longer exists. This clearly highlights that such suspicious domains are actually short lived and created with a dedicated purpose of spamming. To further estimate the average number of times users tried to visit the links coming from such domains, we looked at the cumulated count of corresponding Bitly warning pages. Total number of click requests made to these dead domains only in the month of October was found to be 9,937,250. Spammers thus focus on buying cheap domains to host malicious content and most of these domains eventually die after achieving a good number of hits.
\section{Referrer Analysis}
Next, we identified the top referring domains of these suspicious links. For this, we extracted the link click pattern from referring domain statistics obtained using the Bitly API. Table~\ref{TopReferrers} shows the top 5 domains along with their click frequencies in our dataset. In our results, direct clicks from email clients, messenger, chat applications, or SMS etc. accounted for 20.07\% of all the clicks. Domain \textit{www.hkgolden.com}, a popular computer information forum in Hong Kong was found to generate the maximum number of clicks, 52.84\% of all. In comparison, even the top social networking sites (Facebook and Twitter) only accounted for 5.44\% and 1.87\% of the total traffic respectively. Surprisingly, this behavior is quite different from that observed in the earlier research \cite{3,13,12}, where direct clicks were the maximum and OSM also contributed more than what we found. One possible reason could be that the approach of spammers evolve over time and they always keep changing their focus to the most trending online media. Direct messages have always been targeted by spammers, but here we also identify a drift from OSM to internet forums. This can be due to a large active user base on these discussion forums and comparatively less security features implemented, thus an easy catch for attackers.
\begin{table}[h]
\small
 \begin{center}
    \begin{tabular}{|l|l|}
    \hline
    {\bf Referring Domain}                         & {\bf Clicks}       \\ \hline
        www.hkgolden.com                                    &    453,698,974  \\ \hline
        direct (email, IM, apps)                               &    172,328,021 \\ \hline
        www.facebook.com                     &    46,702,265  \\ \hline
        twitter.com                          &    16,032,903  \\ \hline
        www.poringa.net                      &    9,643,144   \\ \hline
 \end{tabular}
\caption{\label{TopReferrers} Top 5 referring domains for suspicious Bitly links.}
 \end{center}
\end{table}

\begin{figure}[h]
     \begin{center}
		\includegraphics[scale = 0.5]{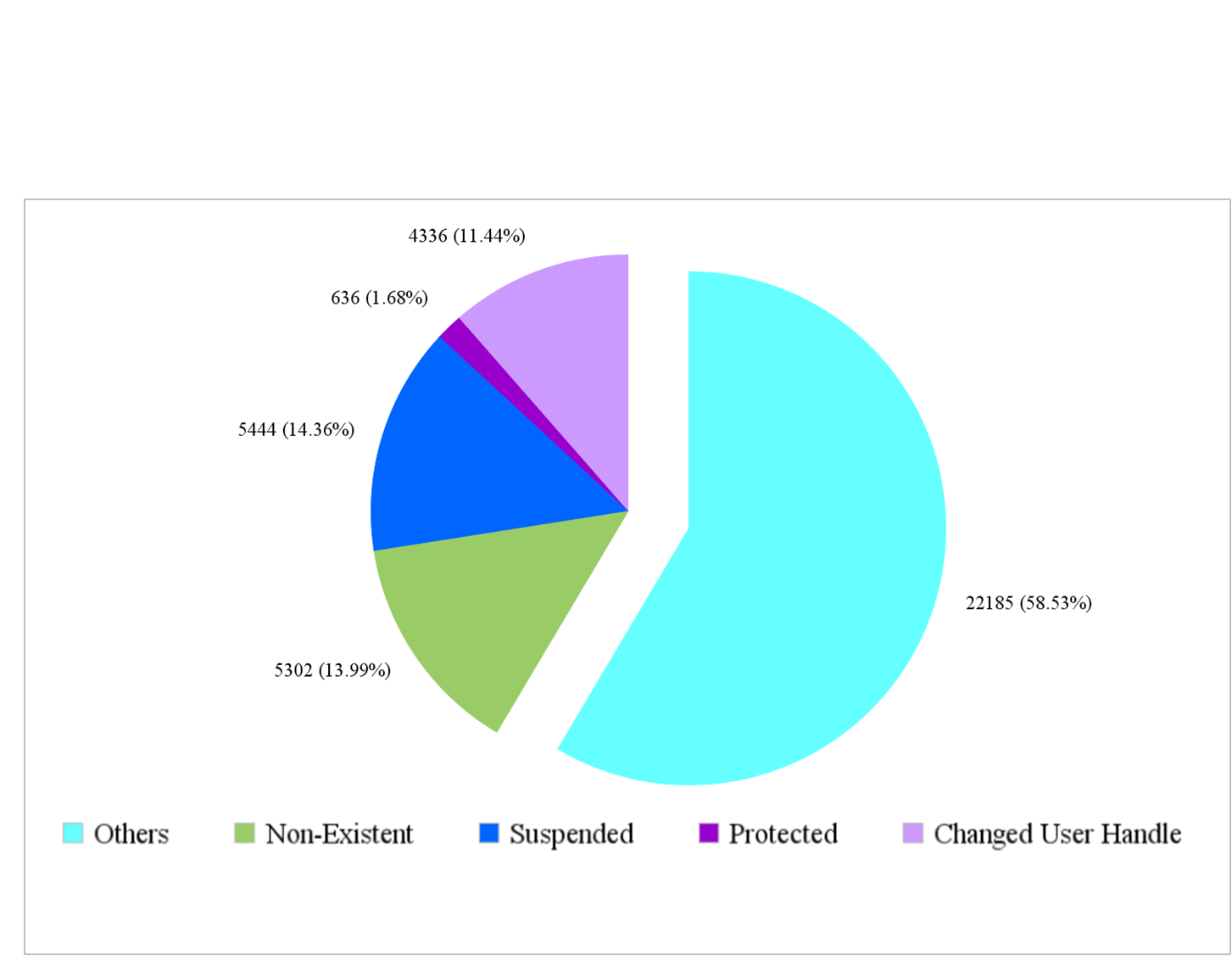}
	\caption{Tweets from referrers.}
		\label{tweetsFromReferrers}
     \end{center}
\end{figure}

In order to understand the characteristics of these referrers, we performed an experiment and limited our study to only Twitter. We did not study Facebook because most of the profiles and posts on Facebook are generally not public. The internet forum we obtained as the top referrer (\textit{www.hkgolden.com}) is also difficult to analyze since it would require web crawling along with language translation. Thus, we extracted Twitter based referrers in our dataset and obtained 37,903 tweets and 21,679 associated Twitter handles. Automated rechecking of all these tweets was also done to determine their state. Doing so, we found that 5,302 (13.99\%) tweets no longer exists and 5,444 (14.36\%) tweets came from 3,729 (17.20\%) users who are now suspended. Fig~\ref{tweetsFromReferrers} gives the corresponding pie-chart representation. Also, 636 of the tweets were protected, which left us with 26,521 tweets and 13,442 Twitter handles that we could actually analyze. We also observed that 3,289 (15.17\%) users changed their Twitter handles during the time of our study. One possible explanation for this behavior could be that these are malicious users who try to change their identities so as to evade or delay detection by Twitter. 

\begin{figure*}[ht]
\begin{center}
       	\subfigure[]{%
            \label{fig:dtitgp2}
            \includegraphics[scale=0.37]{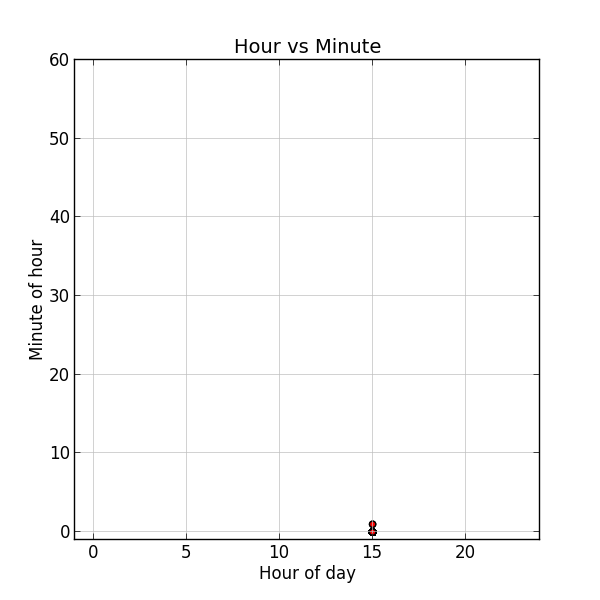}}
\subfigure[]{
    \label{fig:fujisakikaoru}
    \includegraphics[scale=0.37]{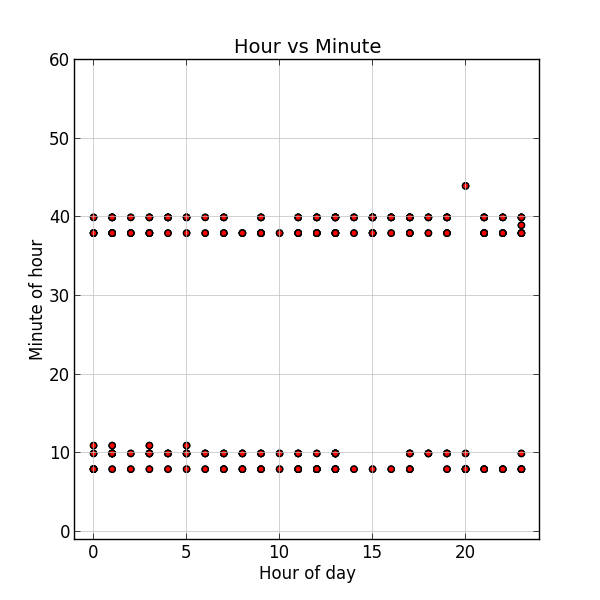}}
\caption{Hour of the day vs. Minute of the hour graph for Twitter user - (a) \textit{@dtitgp2}. (b) \textit{@fujisakikaoru}.}
\end{center}
\end{figure*}

In order to further analyze these Twitter profiles, we extracted their last 200 (or less) tweets using the Twitter API~\footnote{\url{https://dev.twitter.com/docs/api/1.1}}. Here we assume 200 tweets as a reasonable sample to study a user behavior. To examine the presence of Twitter profiles that use Bitly with a dedicated purpose of spamming, we proceeded to identify their tweet pattern. For this, we computed the \textit{Jaccard Similarity Coefficient} over all pair of tweets for each user by removing the URLs if any. \textit{Jaccard Similarity Coefficient} is used to compute the similarity between two sample sets (A and B) and is defined as: 
\begin{equation}\label{eq:jaccard}
J(A,B)=\frac{|A\cap B|}{|A\cup B|}
\end{equation}  
After computing the Jaccard coefficient for each tweet pair from a user, we calculated its overall variance. Users with extremely low variance here signifies the ones who post similar tweet text. In addition to this, cross URL (expanded) and domain level matching was also performed. With this we found 17 Twitter profiles with extremely low variance (\textless 0.00012) on tweet and URL / domain similarity. On a manual annotation of these profiles, we found 3 suspicious profiles which posted only explicit pornographic content and looked like bots and 1 suspicious profile posted links promoting work from home scam. 11 profiles had less than 3 tweets, so we ignored them and 2 profile got suspended. Fig~\ref{fig:dtitgp2} and Fig~\ref{fig:fujisakikaoru} shows hour of the day versus minute of the hour graph for tweets posted by 2 identified suspicious profiles, \textit{@dtitgp2} and \textit{@fujisakikaoru} respectively. The graph clearly shows an automated pattern of posting tweets and confirms that both of these profiles are actually bots. On further inspection we found that all these suspicious profiles are still active and posting malicious links. Since all this was done programmatically and we did not manually annotate each and every account, we might have missed highlighting other suspicious profiles. To verify this, we manually annotated the Twitter profiles which referred a large number (at least 100) of malicious Bitly links from our link-metric dataset. Using this we filtered 23 profiles which are not suspended by Twitter and still actively posting already detected malicious Bitly links. All this makes it evident that the detection of malicious Bitly URLs can also help identify suspicious Twitter profiles. The results also highlights that only detection but no deletion of these Bitly links does not restrict the activities of spammers.

%


\section{Connected Network Impact Analysis}
Bitly allows its users to connect to any number of Facebook and Twitter accounts. With these accounts connected, it becomes easy for a user to shorten and share Bitly links at one click. In this section we present as to how Bitly users abuse this service. To investigate, we first extracted all the encoders~\footnote{We use `encoders' and `users' interchangeably.} of URLs in our link-metric-dataset and found 12,344 distinct Bitly users (encoders) from 413,119 malicious Bitly URLs. Next, we used the Bitly API to collect information about their connected profile network. With this we obtained 5,375 users with a connected Twitter or Facebook account. Looking at individual level, we found 3,415 (63.54\%) users connected only Twitter, 951 (17.69\%) only Facebook, and 1,009 (18.77\%) connected both the networks in our dataset. This gives a clear understanding that majority malicious users prefer to connect their Twitter profiles over Facebook. Possible explanation for low Facebook connections could be because Bitly does not allow a user to connect Facebook brand or fan pages for free.~\footnote{\url{http://support.bitly.com/knowledgebase/articles/76455-how-do-i-connect-my-facebook-and-twitter-account-t}} Limitation to connect to only a personal Facebook account might restrict spammers to disseminate more malicious content in public, but there is no such documented limitation for Twitter. 

On further inspection of the data, we found 507 Bitly users connected multiple Twitter accounts of which 28 users connected at least 10 Twitter accounts. Analyzing these profiles in the same manner we studied the referrers, we extracted their the top (last) 200 tweets. Our target here was to compare multiple profiles connected by each user and infer a possible reason behind their connection. We extracted the URLs posted by all Twitter accounts connected to one Bitly user and did a cross URL / domain comparison. We also collected the link history (last 100) of these users using the Bitly API. All these links were checked for a Bitly warning page by making get requests in python. At last we manually annotated these accounts by looking at the tweet text and URL similarity scores. From this experiment we could identify 3 different cross-network communities that existed across Bitly and Twitter in our dataset -

\begin{figure}[h!]
 	\centering
		\includegraphics[scale = 0.88]{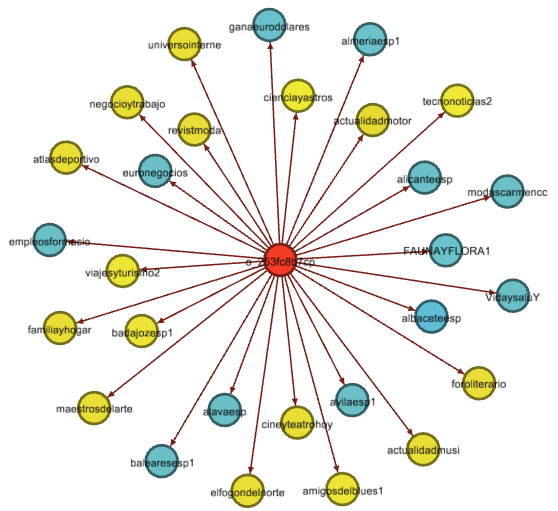}
	\caption{Bitly account from community 2 (\textit{o\_253fc8b7cp}) connected to 28 Twitter accounts. Red node denotes the Bitly account, yellow node represents active Twitter accounts, and blue node represents non-existent Twitter accounts.}
		\label{community2}
\end{figure}

\begin{figure*}[ht]
\centering
       	\subfigure[]{%
            \label{fig:viajesyturismo2}
            \includegraphics[scale=0.37]{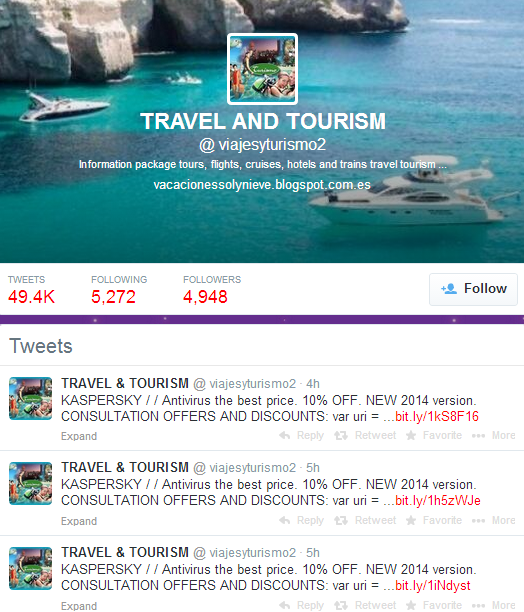}}
\subfigure[]{
    \label{fig:NghiTuOne}
    \includegraphics[scale=0.37]{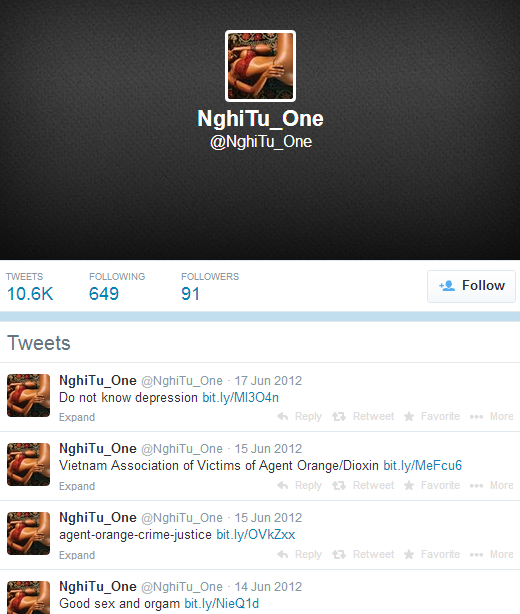}}

\vspace{-7pt}
\caption{Twitter profile snapshot of - (a) User \textit{viajesyturismo2} of community 2. (b) User \textit{NghiTu\_One} of community 3.}

\end{figure*}

\begin{itemize}
 \item {\bf Community 1}:  The first community consisted of 27 Bitly users with 1 associated Twitter account each. All these users shortened links from the same domain and had similar looking user handles starting with ``o\_'' followed by some random hash. Last 100 links shortened by all these users were also redirected to a Bitly warning page. Most surprisingly, all the associated Twitter accounts got suspended when we checked these profiles later and the Bitly profiles looked dormant. It confirms the existence of this malicious community which used Bitly as a medium to propagate spam on Twitter. 
  \item {\bf Community 2}: Another community that we detected consisted of 2 Bitly users with 28 associated Twitter accounts each. Also, during the course of our study one of these Bitly accounts connected 2 more Twitter accounts. 13 of the accounts did not exist when we checked again and other 45 looked malicious and shared similar tweet text and URLs. This community of 2 Bitly and 45 Twitter accounts appears to be an active spam campaign. Fig~\ref{community2} presents the connected Twitter account network corresponding to one Bitly user (\textit{o\_253fc8b7cp}), and Fig~\ref{fig:viajesyturismo2} shows the snapshot of one Twitter profile (\textit{@viajesyturismo2}) that belongs to this campaign. 
y\item {\bf Community 3}: The third community composed of 2 Bitly accounts with 9 Twitter accounts each. All these 18 Twitter accounts shared similar explicit pornographic content. This malicious community is dormant on Bitly but still active on Twitter. Fig~\ref{fig:NghiTuOne} shows the snapshot of Twitter user \textit{@NghiTu\_One}, a part of this community.
\end{itemize}

These communities originated from Bitly to spread malicious content on Twitter. Presence of such big communities propagating bad content clearly highlights the abuse of connected network on Bitly, more importantly because Bitly imposes no limit on the number of networks a user can connect.


\section{System Security Analysis}
Till now, we looked at a generalized characterization of malicious Bitly links by studying the domains they arrive from, the target domains, as well as their connected social network. The above results highlight that how spammers use Bitly as a start point to propagate spam over other media. It now becomes important to comprehend whether Bitly is taking enough measures to deal with such a content and the users it is coming from. In this section, we do a focused study to understand as to how Bitly react in this situation.
\subsection{Efficiency Analysis}
To infer the efficiency of Bitly in detection of malicious links and users, we conduct a two step experiment.
\subsubsection{Comparison Against Some Popular Blacklists}
We performed a check for malicious Bitly links detected by three popular blacklist services- Anti-Phishing Working Group~\footnote{\url{http://www.antiphishing.org/}} (APWG), VirusTotal~\footnote{\url{https://www.virustotal.com/}}, and SURBL~\footnote{\url{http://www.surbl.org/}}. To collect data from APWG, we requested for APWGs live feed service and set it up on our MySql database writing some python scripts. With this service, we collected the data for 6 months (October 2013 - March 2014) and obtained a total of 142,660 and a daily feed of around 746 APWG marked malicious links. Of these, we got 216 Bitly links directly. In order to extract more Bitly links from this dataset, we performed a reverse link lookup by shortening the long malicious URLs and checking their existence on Bitly using Bitly API. Whenever a link is shortened using Bitly API, it returns a parameter called \textit{new\_hash}, indicating whether this link is shortened for the first time or is pre-existent. We collected only the pre-existent links using this technique. With the direct and reverse lookup, we obtained a total of 2,872 APWG marked malicious Bitly links. Next, we wrote a python script to make get requests to Bitly and check if Bitly gives a warning page for these malicious URLs. To our surprise, Bitly could only detect 382 of 2,872 (13.30\%) malicious links. Though Bitly does not claim to use APWG, but such low rate still looks alarming as APWG is a popular and trusted source to detect phishing. 86\% undetected phishing links in 6 months does not give a good picture about the spam detection techniques used by Bitly. 

In parallel to APWG data collection, we also looked for the malicious links marked by VirusTotal over the same time period. VirusTotal is a free online service that characterize a URL as malicious using 52 different website / domain scanning engines and datasets~\footnote{\url{https://www.virustotal.com/en/about/credits/}}. For this, we implemented a web crawler in python and set up a cron job to perform daily look ups on VirusTotal to capture all Bitly links it marks as malicious. With this, we received 569 malicious Bitly links. We again checked these links for a Bitly warning page and found 407 of 569 (71.53\%) links undetected by Bitly. Since VirusTotal is a collated result set from 52 detection services, such high undetection rate again highlights how Bitly misses on a lot of spam.

\begin{figure}[h!]
 	\centering
		\includegraphics[scale = 0.35]{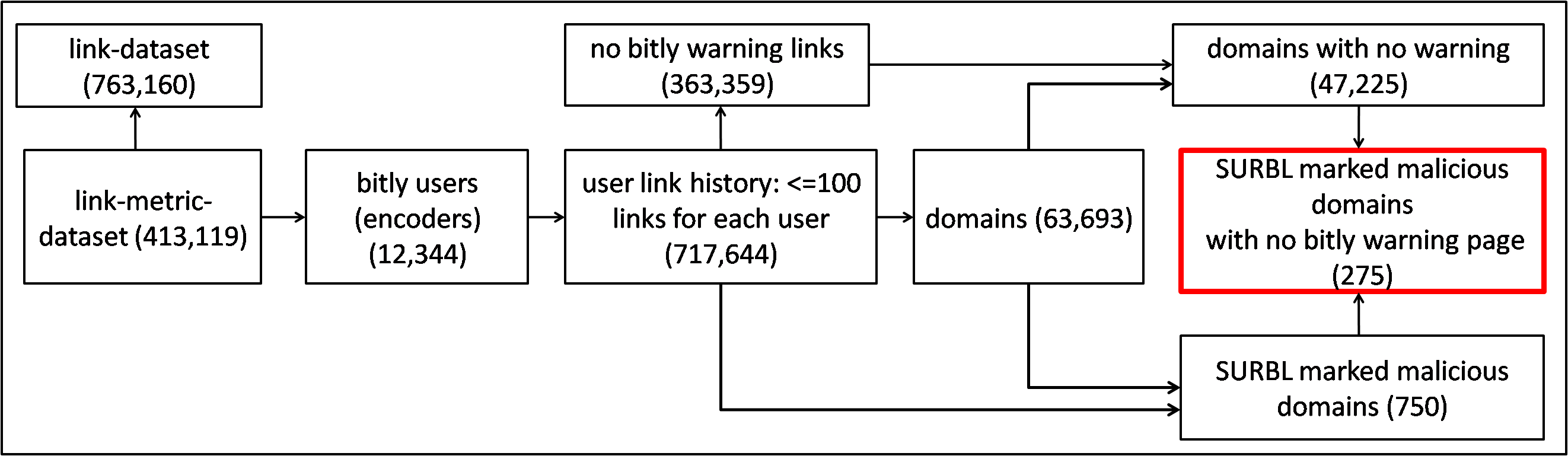}
	\caption{Process diagram to find SURBL blacklisted domains undetected by Bitly.}
		\label{SurblBitlyUndetected}
\end{figure}

In addition to these popular blacklists which Bitly does not claim to use in its spam detection process, we considered a third measure (SURBL) that Bitly profess to apply. SURBL is a domain or IP level blacklist created from the occurrence of websites in unsolicited messages. Fig~\ref{SurblBitlyUndetected} shows the process diagram for this experiment. To check against SURBL, we used Bitly API to collect the link history (last 100 or less) of all encoders of the suspicious Bitly links in our link-metric-dataset. We received 717,644 Bitly links from 12,344 distinct Bitly encoders. On link expansion, the number of associated domains found were 63,693. Links corresponding to 47,225 domains displayed no Bitly warning page. We also checked these domains against SURBL using SURBL API~\footnote{\url{https://pypi.python.org/pypi/surblclient/}} implemented in python and obtained 750 domains blacklisted by SURBL. Looking at the intersection, we found 275 (36.66\%) domains blacklisted by SURBL but undetected by Bitly.
\begin{figure}[h!]
 	\centering
		\includegraphics[scale = 0.53]{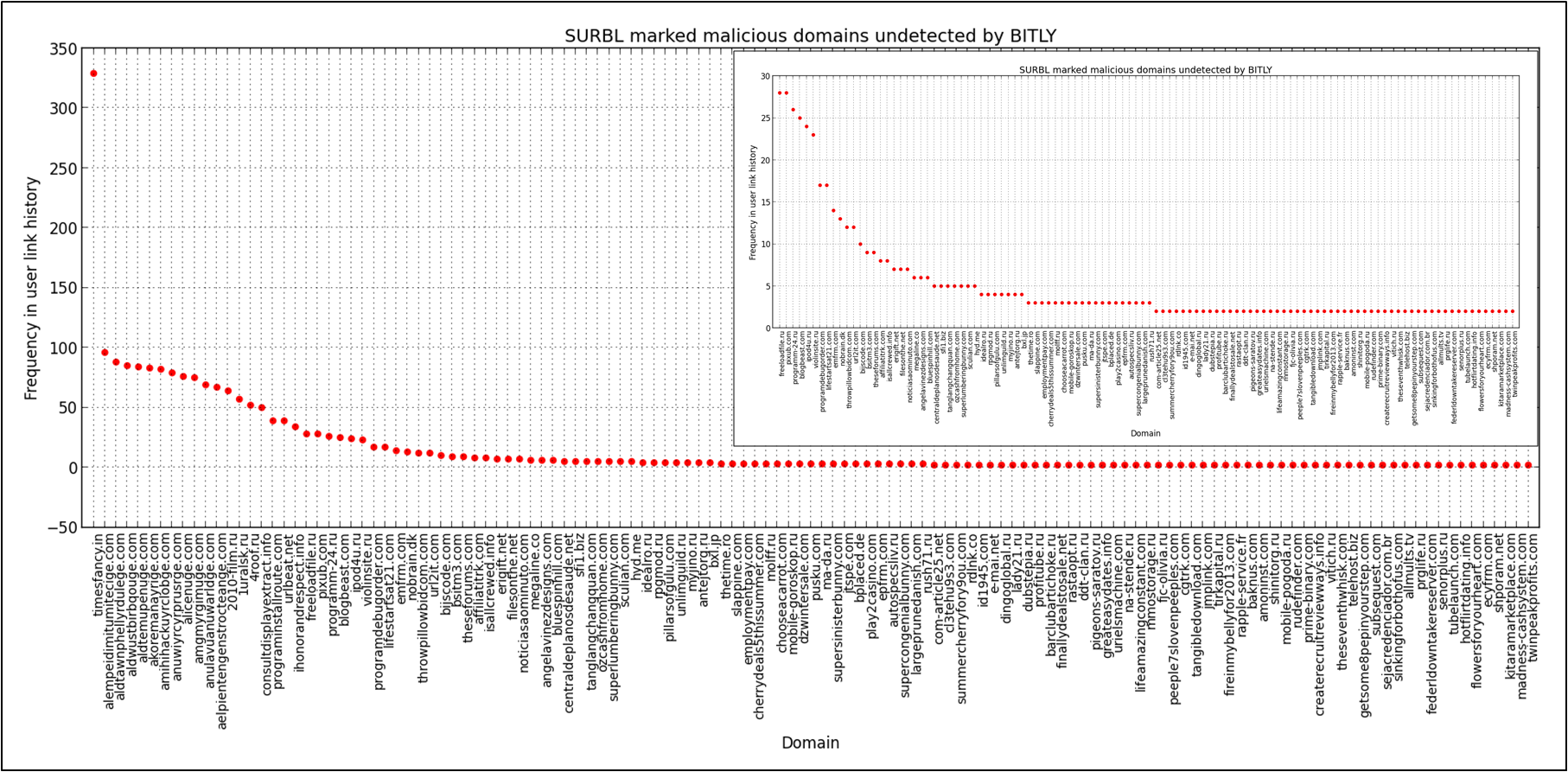}
	\caption{Frequency distribution of SURBL domains undetected by Bitly (with frequency more than 1).}
		\label{surblDomainsUndetected}
\end{figure}

These 275 domains contributed to 2,244 links in our dataset. Fig~\ref{surblDomainsUndetected} presents the frequency distribution of 129 of these undetected domains with occurrence more than one. Subgraph in the same figure starting from domain \textit{freeloadfile.ru} highlights the frequency distribution more clearly. This shows that there are multiple Bitly links corresponding to most of these domains, maximum being 329 for domain \textit{timesfancy.in}. Also, there are 16 domains with frequency more than 50. These experimental results clearly highlights that leaving other popular blacklists aside, Bitly is not even using the claimed spam detection services effectively. Also, such undetected domains contribute to a large number of links if looked at a greater scale. Thus, letting bypass a single malicious domain can act as an invitation to a huge amount of spam.
\subsubsection{Suspicious User Profile Identification}
\begin{figure}[h!]
 	\centering
		\includegraphics[scale = 0.37]{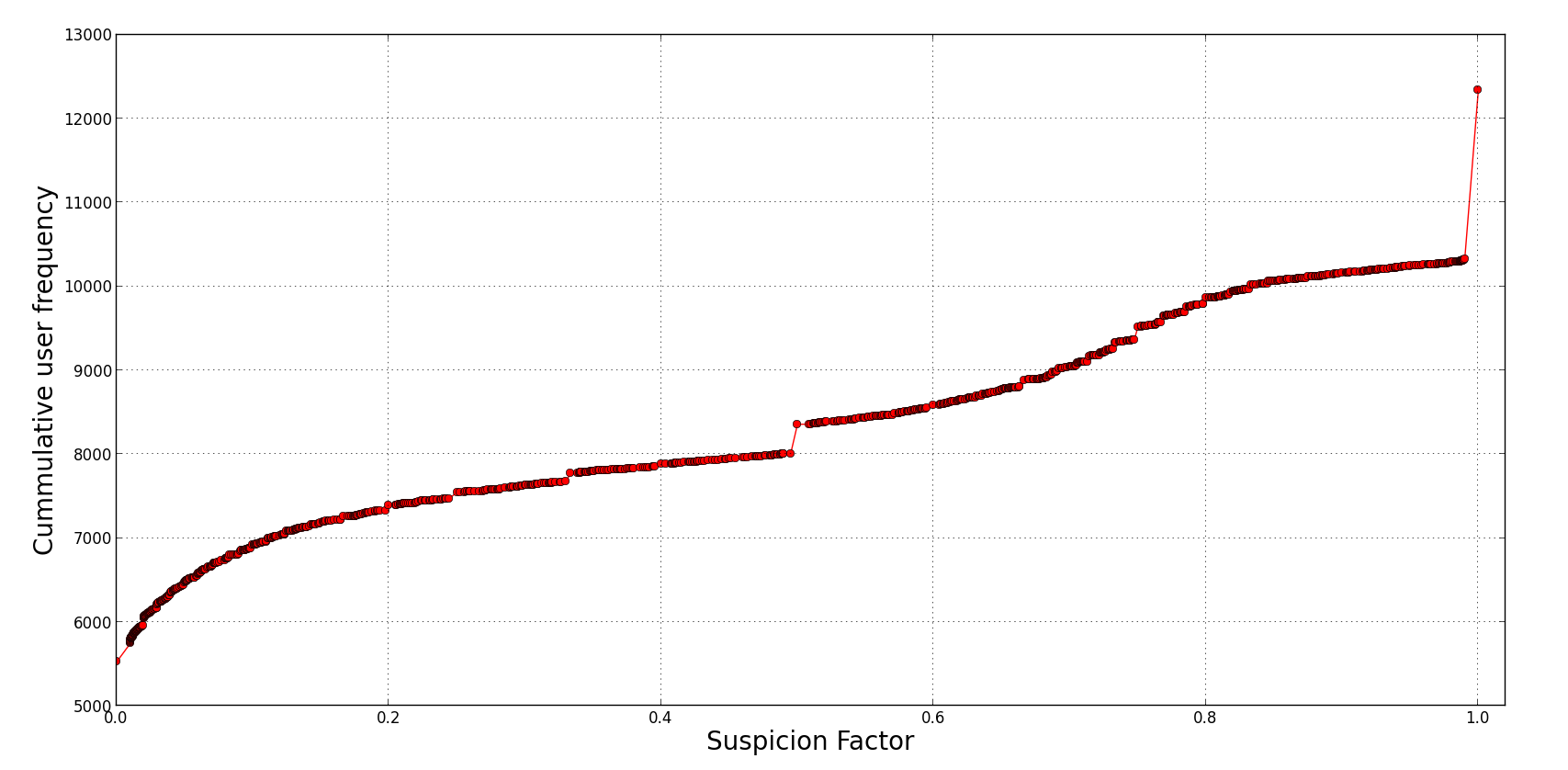}
	\caption{Cum
ulative distribution on number of Bitly users posting suspicious links.}
		\label{cumulative}
\end{figure}

After looking at the inefficiency of Bitly in identifying suspicious links, we proceeded with the detection of suspicious Bitly accounts. On the analysis of all links in encoder's link history, we obtained 112,697 links redirecting to a Bitly warning page (from 12,344 encoders), giving us more suspicious URLs. With this knowledge, we looked at the fraction of suspicious links shortened by different encoders by assigning a \textit{Suspicion Factor} for each as :-\\
\begin{equation}\label{eq:suspicionFactor}
Suspicion~Factor = \frac{\#Links~redirecting~to~Bitly~warning~page}{\#total~links~collected}
\end{equation}  
Here we define Suspicion Factor as the ratio of shortened links giving a Bitly warning page to the total links collected for each encoder. Fig~\ref{cumulative} shows the cumulative distribution on number of Bitly users based on their Suspicion Factor. The graph shows that 12,344 users had a Suspicion Factor less than or equal to 1, and 10,326 users had a Suspicion Factor less than or equal to 0.99. This means that 2,018 (12,344 - 10,326) out of 12,344 encoders (16.35\%) had a Suspicion Factor of 1, indicating that they shortened only suspicious links. Also 2,558 encoders (20.72\%) had at least 80\% of their shortened URLs as malicious (Suspicion Factor \textgreater= 0.8). This clearly highlights the malicious intent of these encoders on creating their Bitly accounts. As of now, Bitly follows a no user suspension policy and does not even delete a malicious link.~\footnote{\url{http://blog.bitly.com/post/138381844/spam-and-malware-protection}} This facilitates the continued existence of a large number of encoders with such evil motives. All these accounts still exists and looks legitimate on viewing their profile. Its only when a user visits multiple links from these profiles leading to malicious content, he gets to know that the profile is created for a dedicated purpose of spamming. This approach is quite different from that followed by Twitter, wherein a suspicious user once detected is immediately suspended to prevent further dissemination of bad content.~\footnote{\url{https://support.twitter.com/articles/18311-the-twitter-rules}} Looking at the extent to which spammers are leveraging the policies used by Bitly, it becomes important for Bitly to also create a workaround to this problem. If not breaking the links, one simple solution could be to assign a credibility score (like Suspicion Factor) to each profile so as to apprise the users visiting that profile of upcoming risks if any. This approach has also been explored by Gupta et al. wherein a tweet can be assigned a credibility score based on various intrinsic characteristics \cite{22}.  
\subsection{Promptness Analysis}
\begin{figure}[h!]
 	\centering
		\includegraphics[scale = 0.75]{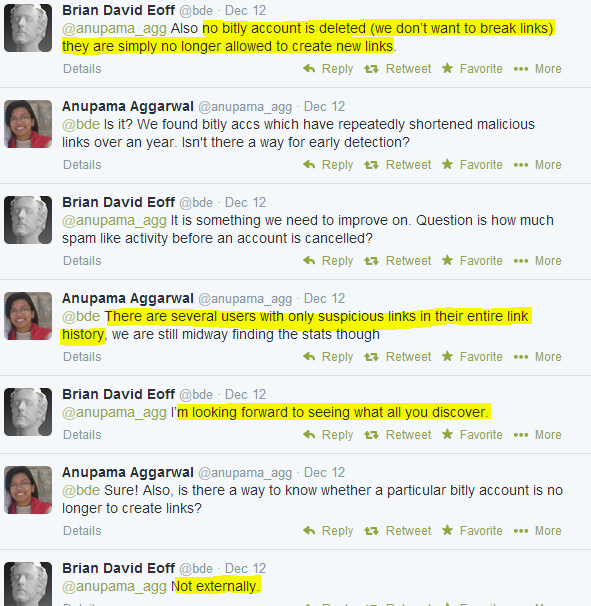}
	\caption{Brian's (Lead Data Scientist at Bitly) reply to out Twitter post about the blog.}
		\label{fig:brianReply}
\end{figure}

\begin{figure*}[h!]
\centering
       	\subfigure[]{%
            \label{fig:bamsesang}
            \includegraphics[scale=0.37]{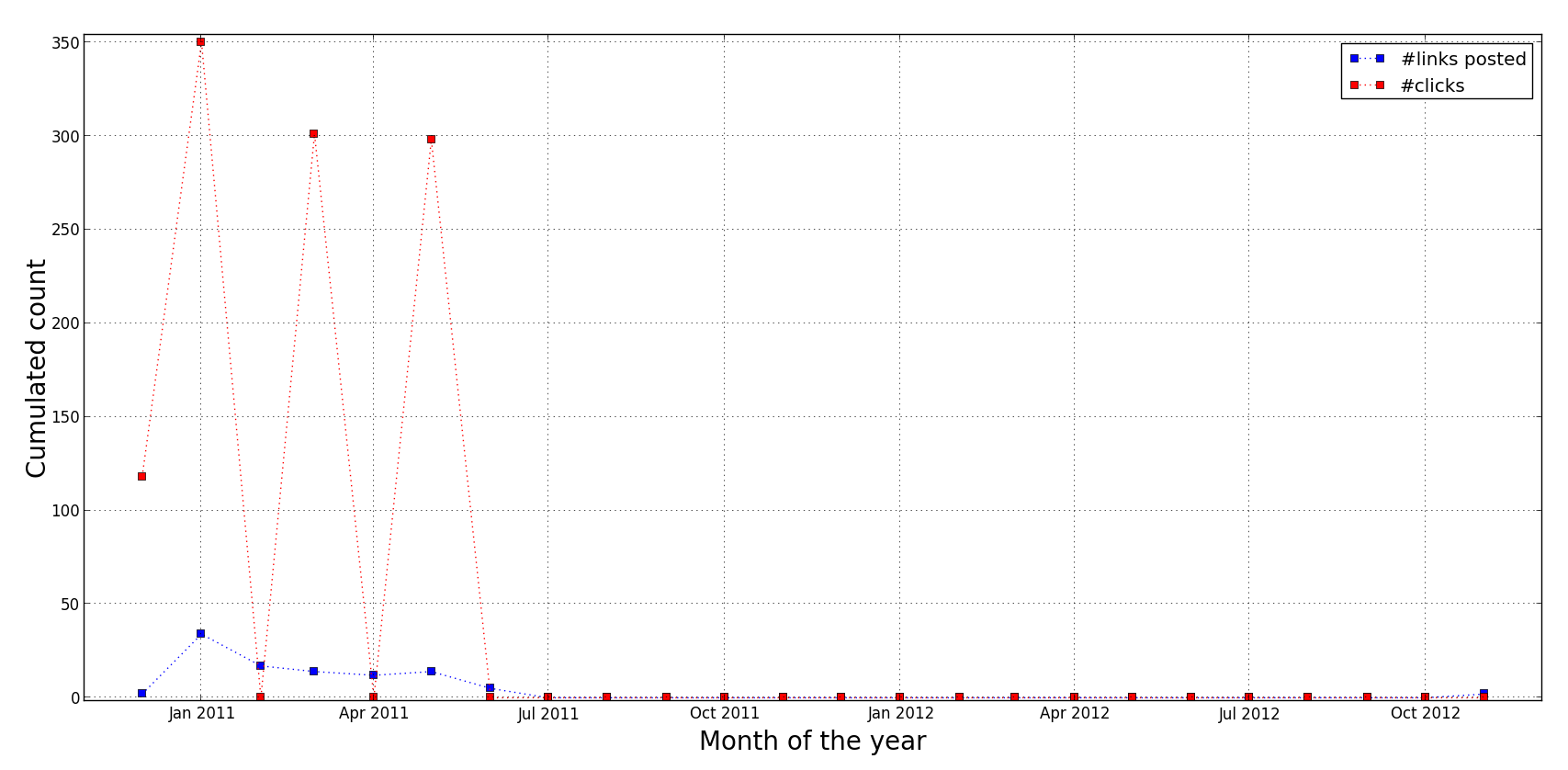}}
\subfigure[]{
    \label{fig:iplayonlinegames}
    \includegraphics[scale=0.37]{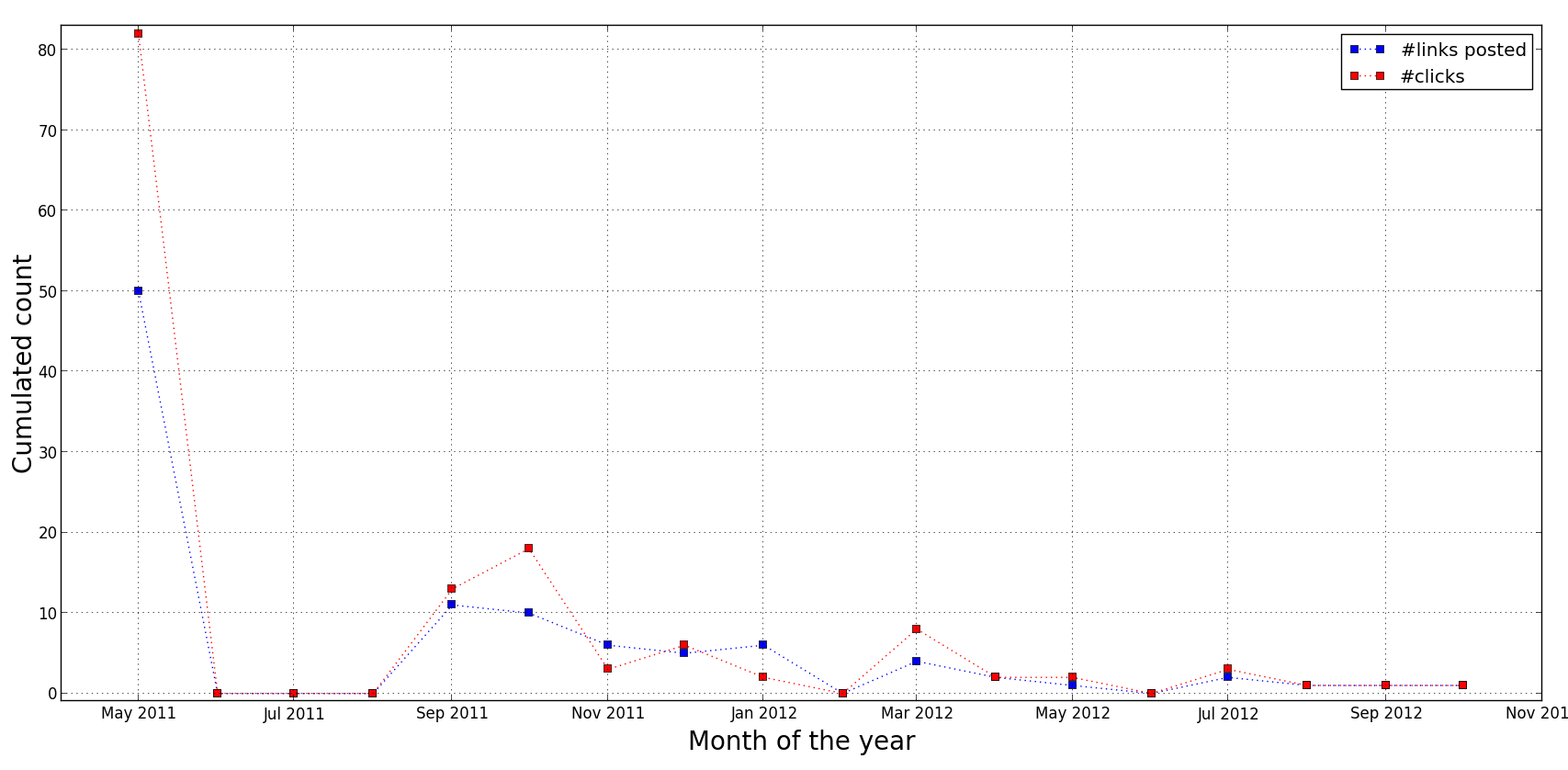}}

\caption{(a) Link history timeline for user \textit{bamsesang}. (b) Link history timeline for user \textit{iplayonlinegames}. The link sharing interval and click pattern clearly reflects the malicious activity being carried out for a long time.}
\end{figure*}

Our results in the above experiment clearly brings to light as to how Bitly users keep shortening only malicious URLs. Till this point of our study, it was unknown to us as to how (if at all) Bitly reacts to suspicious user profiles. To report our findings and get some insights, we made a blog entry on our initial data analysis, in response to which we were informed that Bitly does not suspend user accounts but forbids suspicious users from creating any new links (Fig~\ref{fig:brianReply}).~\footnote{\url{http://precog.iiitd.edu.in/blog/2013/12/bitly-could-do-better/}} Also, since this information is only known to Bitly and the user who is no longer allowed to shorten the links, we could not get this data. But to verify this claim, we performed another experiment to observe the promptness of Bitly in detecting suspicious user profiles. 
For this, we only considered the highly suspicious profiles obtained from the last experiment. We label a profile as highly suspicious if the user has shortened at least 100 links in his link history and all the links give a Bitly warning page, i.e. user has a Suspicion Factor of 1. Using this filter, we obtained 80 such highly suspicious profiles in our dataset. \textit{encoder\_link\_history} metric from the Bitly API was then used to extract the creation time and number of clicks for all 100 links from each profile. At last, we cumulated the number of links created and clicks received per month and formulated a timeline for each user. The maximum month lag of shortening malicious links that we observed was 24 months for user \textit{bamsesang} (Fig~\ref{fig:bamsesang}), followed by 18 months for user \textit{iplayonlinegames} (Fig~\ref{fig:iplayonlinegames}). The timeline shows that user \textit{bamsesang} shortened all malicious links for 7 months, remained inactive for close to one year and then shortened the bad URLs again. On the other hand, user \textit{iplayonlinegames} remained active throughout. These users posted links even when the number of clicks received were less. This highlights that they might be posting links randomly and not monitoring their impact. In contrast, some malicious links shortened by these users did not go unnoticed and received a significant number of hits. Out of 80 highly suspicious users that we labeled, 7 users posted with the month lag greater than 5. These results clearly show an extreme hold-up in suspicious user identification (if at all) by Bitly. 

All this was observed when we only took into consideration past 100 (or less) continuous malicious links for each user. There could have been highly active suspicious users who shortened 100 or more malicious links within a single day or so, but had a very high value of month lag if all their links are studied. Since Bitly API gives only last 100 entries in a single request, this would have required making multiple requests per user to capture the complete link history. We did not do this in our study due to space and time constraints. But in order to check if these highly suspicious users have actually been forbidden by Bitly, we collected their recent link history (after 1 January 2014). We found that 4 of these 80 users were still active and propagating malicious content. Table~\ref{highlySuspiciousUndetected} lists these 4 users along with their last Bitly link posted date. Even for the rest, it can not be said if they have been prohibited from link creation by Bitly or themselves did not create more links. This looks really surprising since Bitly asserted that it disallows suspicious users to shorten more links. All this evidently signifies the ease of penetration of spammers on Bitly and delay in its suspicious user detection process which it actually claims to follow.
\begin{table}[ht]
\small 
\begin{center}
\parbox{.45\linewidth}{
\centering
    \begin{tabular}{|l|l|}
    \hline
    \bf{Bitly user}       & \bf{Date of last link posted} \\ \hline
    insiderslive     & 20 March 2014 \\ \hline
    o\_2flgs1l3eo    & 20 March 2014 \\ \hline
    joeschellenberg4 & 5 March 2014 \\ \hline
    jdkrause         & 15 January 2014 \\ \hline
    \end{tabular}
    \caption {\label{highlySuspiciousUndetected}Confirm undetected highly suspicious Bitly users.}
}
\hfill
\parbox{.45\linewidth}{
\centering
    \begin{tabular}{|l|l|l|}
    \hline
    \bf{Bitly hash} & \bf{\#Warnings} & \bf{Click state} \\ \hline
       14UHah9              &    2,180,862                & no                             \\ \hline
       18PVWD4              &    1,854,529                & no                             \\ \hline
       9eTLS0               &    1,009,871                & yes                            \\ \hline
       18lMytM              &    464,354                 & yes                            \\ \hline
       olZa60               &    289,687                 & yes                            \\ \hline
       13fqZwE              &    109,646                 & yes                            \\ \hline
       4x4Ot5               &    102,698                 & yes                            \\ \hline
       19EH50H              &    88,698                  & no                             \\ \hline
       A2p3qT               &    70,375                  & yes                            \\ \hline
       14VvSWK              &    45,380                  & yes                            \\ \hline
    \end{tabular}
    \caption {\label{highWarningLinks}Top 10 popular malicious Bitly links and their recent click state.}
}
\end{center}
\end{table}

\subsection{Tractability Analysis}
After identifying an extreme delay in the detection of suspicious user profiles by Bitly, we inspected if the access to all popular malicious Bitly links eventually die out after Bitly discovers them. Here we define popular malicious Bitly links as the ones with high number of warning pages displayed. This is important to study because it gives a clear picture about the persistence of already identified malignant~\footnote{We use `malignant' and `malicious' interchangeably.} URL propagation through Bitly network. Also, we restricted our study to only popular links because we wanted to capture the URLs with high overall impact. 

As a sample for this experiment, we extracted the top 1000 Bitly links from our link-dataset based on the number of warning pages reported in the month of October 2013. Bitly API was then used to collect the click history of all these links. Next, we determined and separated the links which got recently clicked (after January 2014). Using this measure, we found 352 out of 1000 malicious links (35.2\%) which are also being actively clicked in year 2014. Table~\ref{highWarningLinks} presents the top 10 of these 1000 popular malicious links along with their recent click state. From the table, 7 of 10 malicious Bitly links with more than 70,000 warning pages already displayed are still active in terms of luring users and receiving clicks. This sample study shows that even though Bitly detected these suspicious links months before, users are still getting trapped and visiting these links. These results help us comprehend that a by-passable Bitly warning page alone is not a strong enough measure to help curtail the dissemination of spam. No control over the access to already detected malicious links can heavily encourage spammers to use Bitly.

\chapter{Data Collection and Labeling for Link Classification}\label{Data Collection and Labeling for Link Classification}
\begin{figure}[h!]
 	\centering
		\includegraphics[scale = 0.6]{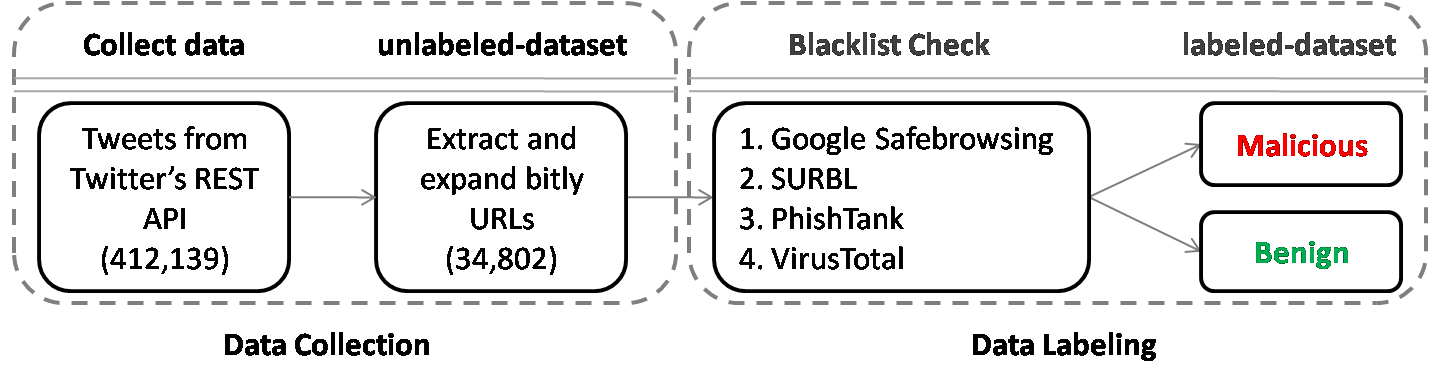}
	\caption{Data collection and labeling.}
		\label{dataCollectionLabeling}
\end{figure}

\onehalfspacing
Now, for malicious short URL detection, we start with data collection for our analysis to build a true positive dataset of malicious Bitly links. This process involves two steps as shown in Fig~\ref{dataCollectionLabeling}, (i) collecting data from Twitter, (ii) labeling the short URLs as malicious or benign.
\section{Data Collection}
In order to collect the data, we used Twitter Rest API~\footnote{\url{https://dev.twitter.com/docs/api/1.1}} and its ``search" method to get only tweets with a Bitly URL. Here we restricted our search query to ``bit.ly" and collected a total of 412,139 tweets with 34,802 distinct Bitly URLs between 12 February 2014 to 15 March 2014. We call this the \textit{unlabeled-dataset}.
\section{Data Labeling}
To formulate a labeled dataset of malicious or benign Bitly URLs, we queried Google Safebrowsing, SURBL, PhishTank, and VirusTotal APIs. Google Safebrowsing~\footnote{\url{https://developers.google.com/safe-browsing/}} is a repository of suspected phishing or malware pages maintained by Google Inc. The Google Safebrowsing API accepts an HTTP GET /POST request to lookup a URL and returns a JSON object describing whether the URL is ``phishing'', ``malware'', or ``ok''.  SURBL~\footnote{\url{http://www.surbl.org/}} is a consolidated list of websites that appear in unsolicited messages. SURBL lookup feature allows a user to check a domain name against the ones blacklisted by SURBL. For this we used SURBL client library implemented in python.~\footnote{\url{https://pypi.python.org/pypi/surblclient/}} PhishTank~\footnote{\url{https://www.phishtank.com/}} is a public crowdsourced database of phishing URLs. The PhishTank API uses an HTTP post request and returns the status of a URL (presence or absence in its database) in a JSON format. VirusTotal~\footnote{\url{https://www.virustotal.com/}} is an aggregated information warehouse of malicious links and domains as marked by 52 website scanning engines and contributed by users. The VirusTotal API also allows an HTTP POST request and gives a JSON response indicating results from all website scanning engines it uses. In order to label our unlabeled-dataset, we mark a Bitly URL as malicious if the corresponding expanded URL or domain is detected by any of these blacklists.  

In addition to these blacklists, we also label a Bitly URL as malicious if it is detected by Bitly itself. Bitly uses various blacklisting services and other measures to detect spam and throws a warning page whenever it identifies a malignant URL. We perform this check for all Bitly URLs in our unlabeled-dataset and label a URL as malicious if a warning page is displayed. Using these techniques, we obtained 8,000 distinct malicious Bitly URLs from our unlabeled-dataset of 34,802 Bitly URLs. We call this the \textit{labeled-dataset}.

\chapter{Feature Selection for Malicious Bitly URL Detection}\label{Feature Selection for Malicious Bitly URL Detection}
\onehalfspacing
Long URL based features to classify a malicious link has been studied over years. Our target in this work is to inspect if short URL based features also hold some distinctive properties to identify a short malicious URL. Since it becomes difficult to capture the intrinsic characteristics of the landing page by using the short URL based features alone, we follow a combined approach by coupling short URL and some long URL based features. This section presents the feature set we identify to classify a Bitly URL as malicious or benign. Table~\ref{featureSet} gives the complete list of features used for our analysis. ~\footnote{We restrict our analysis to only Bitly.}
\begin{table}[h]
\small
\begin{center}
     \begin{tabular}{|p{8cm}|p{8cm}|}
    \hline
    {\bf Feature}                                   & {\bf Description}                                                                          \\ \hline
    \multicolumn{2}{|l|}{\bf WHOIS based features} \\ \hline
    Domain age                                & Difference between domain creation / updation date and expiration date               \\ \hline
    Link Creation domain creation difference  & Difference between domain creation date and Bitly link creation date                 \\ \hline
    \multicolumn{2}{|l|}{\bf Bitly specific features (Non-Click based)} \\ \hline
    Link creation hour                        & Bitly link creation hour                                                             \\ \hline
    Number of encoders                        & Number of Bitly users who encoded a particular link                                  \\ \hline
    Type of encoders           & Ratio of anonymous or application based encoders to total encoders                   \\ \hline
    \multicolumn{2}{|l|}{\bf Bitly specific features (Click based)} \\ \hline
    Link creation-click lag      & Difference in days between Bitly link creation date and date of first click received \\ \hline
    Type of referring domains       & Ratio of referring domains from a direct source to the total referring domains       \\ \hline
    \end{tabular}
\caption {\label{featureSet}Features used for classification of malicious Bitly URLs.}
\end{center}
\end{table}
\section{WHOIS based features}
WHOIS is a query and response protocol that gives information like domain name, domain creation / updation date and domain expiration date for a particular URL. This information is particularly useful to detect domains which are intentionally created for malicious purposes. Most spammers prefer to register their domains for a short duration and also change the domains frequently to evade detection. In our domain analysis section (Chapter~\ref{chapter:Experiments and Analysis}), we observed 83.06\% malicious domains to be non-existent when we rechecked them after 5 months. This gives a clear understanding that malicious domains are usually short lived. Thus, we use WHOIS based features like domain creation and expiration dates to track the lifetime of URLs in our labeled-dataset. Also, it is commonly observed that suspicious domains are created or updated just before they are actually used. Therefore, we also use the difference between domain creation and Bitly link creation time as one of our feature.
\section{BITLY specific features}
Bitly provides a detailed analytics about each link it shortens. This includes some general link properties along with the click and user history corresponding to each Bitly URL. We assume that these analytics contain a lot of hidden properties that can help segregate malicious and benign links.
To test our assumption, we identify some Bitly specific short URL based features and divide these as \textit{Non-Click based} and \textit{Click based}. \textit{Non-Click based} features are the ones which define general characteristics of a short Bitly URL and are independent of its click history. \textit{Click based} features on the other hand depends on the click analytics of a Bitly link.
\subsection{Non-Click based features}
We determine 3 \textit{Non-Click based} features:
\begin{itemize}
 \item {\bf Link creation hour}: Malicious users many-a-times rely on automated mechanisms to shorten and share the links. Link shortening timestamp patterns in case of such automated approaches might not be similar to the genuine usage trend. For example, a malicious user would be more likely to create a link at 0300 hours (IST) than a human. The idea could be to automatically shorten multiple bad URLs late at night to propagate it the next day. This kind of automated behavior can thus be captured by tracking link creation hour of the Bitly links.
 \item {\bf Number of encoders}: Encoders are the list of Bitly users who encode the same Bitly URL. Bitly creates different local hashes but the global hash for all Bitly URLs redirecting to the same page is unique. Number of distinct encoders corresponding to a Bitly URL depicts its popularity. As highlighted in section 5.3, malicious communities take advantage of this feature by creating multiple identities to shorten the same link. This trick helps to spread malicious content at a greater pace. We use this feature in order to detect the presence of such suspicious communities in our dataset.
 \item {\bf Type of encoders}: Encoders can either be regular Bitly users or users who use some third party services provided by Bitly. These third party services like TweetDeck API, Twitterfeed, Tweetbot, etc. provides a single interface for users to shorten and share the links on multiple OSM.~\footnote{\url{https://bitly.com/pages/partners}} Since we collect our dataset using Twitter, we focus on only Twitter based applications. Twitterfeed is a service to feed content to Twitter. TweetDeck API is an API service for TweetDeck, which provides a dashboard to manage Twitter. Tweetbot is a full featured Twitter client for iPhone. Offenders many a times use such services to make their task of sharing malicious links easy. In addition to these third party services, various Bitly links are also shortened anonymously by users. In case of such anonymous shortening, Bitly gives the encoder information as ``someone'' or ``anonymous''. In our link-metric-dataset, we observe traces of malicious users who hide their identities and shorten bad links. Anonymous Bitly link shortening and third party service usage can therefore be used as a feature to identify malicious links.
\end{itemize}
 \subsection{Click based features}
We determine 2 \textit{Click based} features:
\begin{itemize}
\item {\bf Link creation-click lag}: Difference in days between link creation and first click received by a Bitly URL is another distinct feature we recognized. Antoniades et al. stated that most legitimate short URLs are clicked on the same day they are created \cite{3}. The average lifetime of malicious short URLs have also been reported to be higher than that of the legitimate ones \cite{4}. This gives a notion that malicious short URLs do not gain immediate popularity and the number of clicks on such URLs evolve slowly. A possible reason behind this phenomenon can be that malicious URLs are generally shared by less reliable sources which may not receive instant clicks. Hence, we include link creation-click lag as a characteristic feature to capture how quick the short URL resolves.
\item {\bf Type of referring domains}: Domains referring click traffic to a particular link also holds some distinguishing properties to classify a short malicious URL. Our analysis in Chapter~\ref{chapter:Experiments and Analysis} reveals that a large fraction of malicious Bitly URLs get clicked directly through email clients, messengers, chat applications, SMS, etc. To further test these results, we use the fraction of referring domains that contributes to direct clicks as another feature in our classification. 
\end{itemize}

\chapter{Experimental Setup}\label{Experimental Setup}
\onehalfspacing
In this section, we describe the mechanisms used in our classification of malicious short URLs. Our target is to be able to detect such URLs with high accuracy. The experiments involve a 3 step process – i) creating a labeled dataset, ii) training the suitable machine learning classifier, and iii) testing an unlabeled dataset on the trained classifier.  We collected our initial dataset of Bitly URLs using Twitter and labeled it as malicious / benign by checking against popular blacklists, as already discussed. Classification algorithms involve training on such pre-labeled dataset, based on which it predicts the labels of an unseen data. Here we want our classifier to predict if a given unseen Bitly URL is malicious or benign. This section explains the machine learning classifiers we used in our experiments.
\section{Machine Learning Classification}
\label{sec:classifiers}
In order to assess the most appropriate and efficient mechanism to detect malicious short URLs, we inspect various machine learning classifiers which are best suited for our study. For this, we use the classifiers implemented in Weka software package \cite{15}. Weka is an open source collection of machine learning classifiers for data mining tasks. We performed our experiments on some of the popular classification algorithms in Weka. Now, we give a brief description about these classifiers.
\subsection{Naive Bayes Classifier}
This is a simple probabilistic classifier based on the Bayes' theorem. It assumes all classification features to be independent of one another and works best when the dimensionality of inputs is very high. Parameter estimation in naive bayes uses maximum likelihood principle, thus a data point is classified into the class with highest probability. An advantage of using this model is that it does not have a large training data requirement for parameter estimation and classification. It uses variance of variables in each class and is also not sensitive to irrelevant features. To implement this algorithm, we use NaiveBayes classifier in Weka.~\footnote{\url{http://weka.sourceforge.net/doc.dev/weka/classifiers/bayes/NaiveBayes.html}}
\subsection{Decision Tree Classifier}
This is one of the most popular classification techniques that uses decision tree as a predictive model. It follows a rule based approach to observe data features and make conclusions about the item's target value. Decision Tree starts at the root and makes binary (yes / no) decisions at each level until it reaches the leaf node. We use Weka's j48 tree classifier for this study.~\footnote{\url{http://weka.sourceforge.net/doc.dev/weka/classifiers/trees/J48.html}}

\subsection{Random Forest Classifier}
Random Forest is a classifier that consists of multiple decision trees and outputs the class which is the mode of the classes output by individual trees. For each data point, it randomly chooses a set of features for classification. It uses averaging to select most important features hence improve the predictive accuracy and control over-fitting. It can run efficiently on large databases. This method is also useful for estimating missing data and maintains accuracy even when a large proportion of the data are missing. To implement this, we use RandomForest tree classifier in Weka.~\footnote{\url{http://weka.sourceforge.net/doc.dev/weka/classifiers/trees/RandomForest.html}}

\section{Training and Testing Data}
We divide our labeled-dataset into training (75\%) and test data (25\%). Ten fold cross validation is applied on the training data over all classifiers. Here the data is partitioned into 10 subsets. In each test run, 9 subsets are used for training and 1 is used for testing. This training-testing process repeats 10 times and average accuracy of each classifier is analyzed. The remainder 25\% test data is then supplied to determine the actual performance of the classifier on an unseen (unlabeled) dataset, which gives our final classification results. 
 
\chapter{Classifier Evaluation}\label{Classifier Evaluation}
\onehalfspacing
We already defined the feature set and machine learning algorithms used in our classification. This section describes the evaluation metrics and results we obtained from our experiments based on the ground truth dataset.
\section{Evaluation Metrics}
\begin{figure}[h!]
 	\centering
		\includegraphics[scale = 0.5]{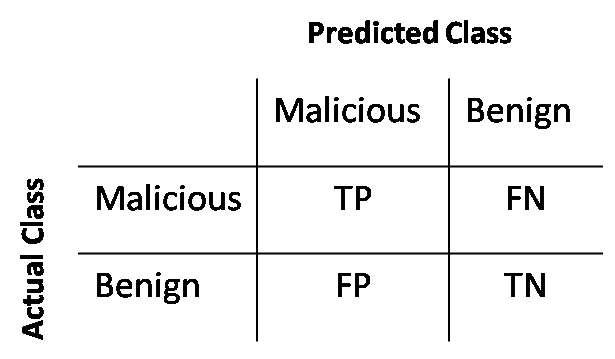}
	\caption{Confusion matrix showing relation between True Positives (TP), False Positives (FP), True Negatives (TN), and False Negatives (FN).}
		\label{confusionMatrix}
\end{figure}

Two major metrics that we use for the evaluation of our classifier are F-measure (FM) and accuracy (A). F-measure is defined in terms of precision (P) and recall (R). Precision (Equation~\ref{eq:precision}) is the proportion of predicted positives in the class that are actually positive, while recall (Equation~\ref{eq:recall}) is the proportion of the actual positives in the class which are predicted positive. This dependency can be better described by a confusion matrix as presented in Fig~\ref{confusionMatrix}. 
Each row of the confusion matrix represents the instances in an actual class, whereas each column represents the instances in a predicted class. Confusion matrix is very useful in understanding the relation between true positives (correctly identified values), false positives (incorrectly identified values), true negatives (correctly rejected values), and false negatives (incorrectly rejected values). With this knowledge, F-measure and accuracy can be easily computed. F-measure (Equation~\ref{eq:F-measure}) is a weighted average of precision and recall, and accuracy (Equation~\ref{eq:accuracy}) is the closeness of measurements to the actual (true) value. 

\begin{equation}\label{eq:precision}
P = TP/(TP+FP) 
\end{equation}
\begin{equation}\label{eq:recall}
R = TP(TP+FN)
\end{equation}
\begin{equation}\label{eq:F-measure}
FM = 2*(P*R)/(P+R)
\end{equation}
\begin{equation}\label{eq:accuracy}
A = (TP+TN)/(TP+TN+FP+FN)
\end{equation}
\section{Evaluation Results}
We now describe the results obtained from our classification experiments. We trained and tested three classifiers - Naive Bayes, Decision Tree, and Random Forest (as discussed in Chapter~\ref{Experimental Setup}) using 7 features described in chapter~\ref{Feature Selection for Malicious Bitly URL Detection}. Our true positive dataset is split into 75\% training and 25\% test set. With this, the classifiers are trained on a ground truth of 5,926 malicious and 6,074 benign short URLs and 10 fold cross validation is applied. At last, we evaluate our classifier on the remainder 25\% dataset (2,074 malicious and 1,926 benign Bitly URLs). Table~\ref{classifierAllFeatures} presents the outcome from each of these classifiers. 
\begin{table}[h]
\small
\begin{center}
    \begin{tabular}{|l|l|l|l|}
    \hline
    {\bf Evaluation Metric}     & {\bf Naive Bayes} & {\bf Decision Tree} & {\bf Random Forest} \\ \hline
    Accuracy              & 72.15\%     & 78.37\%       & 80.43\%       \\ \hline
    Recall (malicious)    & 73.10\%     & 82.40\%       & 81.00\%       \\ \hline
    Recall (Benign)       & 71.10\%     & 74.10\%       & 79.90\%       \\ \hline
    Precision (malicious) & 73.10\%     & 77.40\%       & 81.20\%       \\ \hline
    Precision (Benign)    & 71.10\%     & 79.60\%       & 79.60\%       \\ \hline
    F-measure (malicious) & 73.10\%     & 76.70\%       & 81.10\%       \\ \hline
    F-measure (benign)      & 71.10\%     & 76.70\%       & 79.70\%       \\ \hline
    \end{tabular}
    \caption {\label{classifierAllFeatures} Results of classification of malicious Bitly URLs using all 7 features.}
\end{center}
\end{table}
We notice that as we move from Naive Bayes to Decision Tree, the accuracy and F-measure increases for malicious as well as benign class. With the implementation of Random Forest, both these metrics increases even further. This increase in accuracy by Random Forest is attributed to the point that it considerably reduces the false positives in our classification. Hence, we achieved an overall accuracy of 80.43\% and weighted average of F-measure as 80.40\%. 

We understand that our training dataset is a mix of Bitly links that receive and do not receive clicks. Of the 8,000 malicious ground truth links in our labeled-dataset, we found that 3,693 (46.16\%) links were never clicked by anyone. Although this property itself serves as a feature in the above classification, but we believe that our classifier should perform even better if we segregate all links with zero clicks. This is because such links can be easily classified using only Non-Click based features. Therefore, we separate Bitly links with zero clicks and obtain a true positive malicious dataset of 3,693 (46.16\%) from a total of 8,000 links. We also randomly select 3,693 benign links with no clicks from the labeled-dataset and perform our classification experiments again by removing Click-Based features. Table~\ref{classifierNonClickFeatures} gives the results from this classification. As expected, an increase in the overall accuracy and F-measure for each classifier is observed. With Random Forest, we achieve a high accuracy of 86.41\% and weighted average of F-measure as 86.40\%. Also, when we tested our complete labeled-dataset again by excluding Click based features, we attained a maximum accuracy of 83.50\% and F-measure for malicious class increased to 84.1\% (using Random Forest).

\begin{figure*}[ht]
\centering
       	\subfigure[]{%
            \includegraphics[scale=0.5]{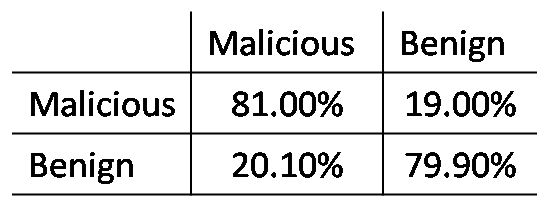}}
\subfigure[]{
    \includegraphics[scale=0.5]{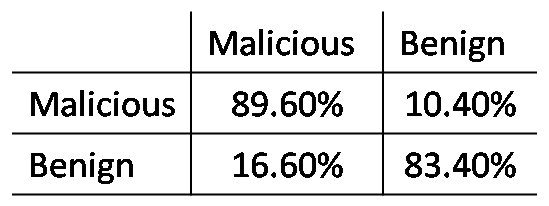}}
\subfigure[]{
    \includegraphics[scale=0.5]{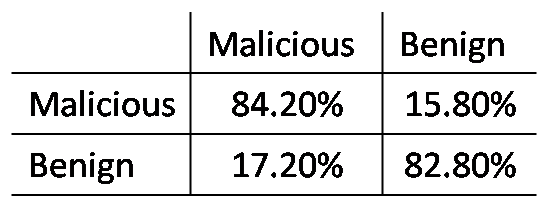}}

\caption{\label{fig:confusionMatrixAllMain} Confusion matrix for - (a) complete labeled-dataset. (b) only Non-Click data from labeled-dataset. (c) complete labeled-dataset by applying only Non-Click based features.}

\end{figure*}

Previous studies on phishing emails and links on Twitter also shows that Random Forest works the best as compared to other classifiers \cite{9,17}. In order to further inspect the performance of Random Forest in our experiments, we analyze the confusion matrix for complete as well as Non-Click labeled-dataset (Fig~\ref{fig:confusionMatrixAllMain}). For complete labeled-dataset, we could correctly identify 81\% of malicious short URLs but 19\% of malicious links were misclassified as benign. On a manual inspection of some of these misclassified profiles, we observed their click pattern captured in our feature set to be quite similar to that of the benign profiles. In case of Non-Click labeled-dataset, the correct classification rate reached 89.60\% and misclassification rate dropped to almost half (10.40\%). Also, on removing Click-Based features from complete labeled-dataset, true positive increased to 84.20\% and false negative decreased to 15.8\%. This indicates that our Non-Click based features are alone good at classifying a malicious Bitly link, irrespective of looking at its click history. Although our proposal is capable of capturing the click patterns of a link, it can be particularly helpful to detect malicious Bitly links at zero hour, i.e. even before it is actually clicked. 


\begin{table}[h]
\small
\begin{center}
    \begin{tabular}{|l|l|l|l|}
    \hline
    {\bf Evaluation Metric}     & {\bf Naive Bayes} & {\bf Decision Tree} & {\bf Random Forest} \\ \hline
    Accuracy              & 80.02\%     & 85.06\%       & 86.41\%       \\ \hline
    Recall (malicious)    & 79.60\%     & 89.50\%       & 89.60\%       \\ \hline
    Recall (Benign)       & 80.40\%     & 80.80\%       & 83.40\%       \\ \hline
    Precision (malicious) & 79.30\%     & 81.50\%       & 83.60\%       \\ \hline
    Precision (Benign)    & 80.70\%     & 89.10\%       & 89.50\%       \\ \hline
    F-measure (malicious) & 79.50\%     & 85.30\%       & 86.50\%       \\ \hline
    F-measure (benign)    & 80.50\%     & 84.80\%       & 86.30\%       \\ \hline
    \end{tabular}
    \caption {\label{classifierNonClickFeatures} Results of classification of malicious Bitly URLs using only Non-Click based features.}
\end{center}
\end{table}

\section{Feature Ranks}
Some features in machine learning classification are more important and useful than others. In this section, we describe the most informative features based on their ranks. We obtain this ranking using Weka's \textit{InfoGainAttributeEval} package for attribute selection.~\footnote{\url{http://weka.sourceforge.net/doc.dev/weka/attributeSelection/InfoGainAttributeEval.html}} This technique evaluates the worth of an attribute by measuring the information gain with respect to the class. Fig~\ref{tb:featureRank} gives the ranked feature lists for classification experiments on complete and Non-Click dataset. 
\begin{figure*}[h]
\small
\centering
       	\subfigure[]{%
\begin{tabular}{|p{1cm}|p{4cm}|}\hline
\small
\textbf{Rank} & \textbf{Feature}\\ \hline
1 & Type of referring domains\\ \hline
2 & Link Creation domain creation difference\\ \hline
3 & Domain age\\ \hline
4 & Link creation hour\\ \hline
5 & Type of encoders \\ \hline
6 & Link creation-click lag \\ \hline
7 & Number of encoders \\ \hline
\end{tabular}
}
\subfigure[]{
\begin{tabular}{|p{1cm}|p{4cm}|}\hline
\small
\textbf{Rank} & \textbf{Feature}\\ \hline
1 & Link creation hour\\ \hline
2 & Link Creation domain creation difference\\ \hline
3 & Domain age \\ \hline
4 & Type of encoders \\ \hline
5 & Number of encoders\\ \hline
\end{tabular}
}
\caption{Feature rank based on information gain for classification on - (a) complete labeled-dataset. (b) Non-Click labeled-dataset.}
\label{tb:featureRank}
\end{figure*}
On a mix dataset of clicked and non-clicked links, \textit{Type of referring domains} is found to be the most discriminating feature. Malicious Bitly links usually have more direct referrers, i.e. such links are propagated more through email clients, mobile / chat applications, SMS, etc. Such sources are often targeted by spammers because they have a direct impact on the user, which increases the chances for a malicious link to be seen and visited. \textit{Link Creation domain creation difference} is another important feature based on information gain. As seen in Chapter~\ref{chapter:Experiments and Analysis}, offenders buy cheap domains for a deliberate purpose of spamming. Most of the offenders do not wait and immediately start using these domains by shortening Bitly links. Since these domains are mostly registered for a short time span, the objective of spammers is to fully exploit them before they expire. The third most informative feature \textit{Domain age} highlights the same. 

Another strategy adopted by spammers is to shorten links at odd hours of the day, as depicted by the fourth most informative feature \textit{Link creation hour}. Such malicious users actually use some automated scripts (bots) to bulk shorten Bitly links at odd hours. These links are then shared on different sources at the time they expect targeted users to read. Other than these top 4 features, \textit{Type of encoders}, \textit{Link creation-click lag}, and \textit{Number of encoders} are found to be comparatively less informative. This implies that these three features did not help much in the classification. In case of classification using our Non-Click dataset, \textit{Link creation hour} is observed to be the most important feature. Remaining features follow the same ranking pattern as in the complete dataset.

%
\chapter{Conclusions, Limitation and Future Work}\label{Conclusion, Limitation and Future Work}
\onehalfspacing
In the first part of our study, we presented an overview of the general characteristics of malicious Bitly links and their propagation on OSM. Spammers shorten malicious Bitly links by buying cheap domains and registering them for a short period of time. We found 83.06\% malicious domain in our link-dataset that received 9,937,250 clicks on Bitly to be non-existent when we rechecked after 5 months. Such domains are created for a dedicated purpose of spamming and eventually die out after targeting a significant number of victims. In our referrer analysis, we observed 52.84\% Bitly click traffic to originate from a popular Hong Kong based discussion forum, and 20.07\% came from direct clicks. This trend is different from what has been generally reported by the previous studies, where direct clicks contribute to the majority.  Shift in the approach of malicious users can be attributed to the large active user base and comparatively less security mechanisms implemented on such portals. We also found that spammers exploit Bitly's policy of not imposing a cap on the number of connected OSM (Facebook / Twitter) accounts. We traced such spammers and detected three malicious communities which operate across Bitly and Twitter. In total, these communities composed of 31 Bitly accounts and 90 connected Twitter accounts spreading spam, scam, or explicit pornographic content.

In the second part of our study, we unveiled some loopholes in the security policies and mechanisms used by Bitly.  It was found that Bitly could not effectively detect malicious URLs already tracked by the popular blacklist services like APWG and VirusTotal. We also observed that Bitly could not detect 36.66\% links identified by a domain blacklist called SURBL, which it actually claims to use. This highlights the inefficiency of Bitly in using even the claimed spam detection services effectively. We also detected significant traces of malicious users who abuse Bitly's no account suspension policy. 16.35\% malicious users in our dataset kept shortening only suspicious links in their history without being noticed. Also Bitly asserts that it forbids such suspicious users from shortening more links, but we identified users who kept shortening bad links for close to two years and their state of detection is still unknown. Finally, we also traced highly popular Bitly links (with large number of warning pages displayed) and found 35.2\% of them being clicked till date. This brings to light that a by-passable Bitly warning page is alone not enough to curtail the overall problem of spam.

In the last part of our study, we proposed a mechanism to detect malicious links on Bitly. We identified click based and non-click based features from Bitly and coupled them with some domain specific features to classify a Bitly URL as malicious or benign. Computation of all these features required close to 50 seconds per Bitly URL (Intel(R) Xeon(R) CPU E5-2640 0 @ 2.50GHz). Our classifier predicted malicious links with the best accuracy of 80.40\% on a mix dataset of clicked and non-clicked links. By only considering non-click based features, our classifier attained an improved accuracy of 83.51\% on the mix dataset, and 86.41\% on only non-click dataset. Thus, our algorithm is not only efficient in detecting malicious Bitly links when they receive user clicks, but can also identify suspicious links at zero hour, i.e. when no click is received. 

We performed our initial analysis on a limited dataset acquired from Bitly. This dataset captured only links detected malicious by Bitly in the month of October 2013. Since the characteristics of spammers change over time, we can do a detailed and comparative analysis on a more exhaustive dataset. Also, Bitly claims to forbid certain suspicious accounts but does not give this information through its API. This created a trouble in interpreting if a dormant but highly suspicious account in our dataset has been blocked by Bitly or not shortening the links itself. 

For malicious Bitly link detection, our classifier currently works better on non-click dataset. This is because we used only 2 click-based features in our classification. In future, we would also like to study as to how click patterns on a malicious Bitly URL evolve over time. This would require a temporal analysis and can serve as another good feature for our classifier. In order to evade on submission check by the URL shorteners, malicious users many a times shorten an already shortened link. This multi-level link obfuscation can therefore be used as another important feature. We can also include the user profile attributes like suspicion-factor for each encoder to achieve better results. This would require extracting encoder information and analyzing their click history for each link. We understand that such an evaluation is a time consuming process. Therefore, we did not include these parameters and performed our current study on only some lightweight and easily computable features. As a part of our future work, we would like to include these parameters since incorporation of such discriminating features can help improve the overall performance of our classifiers. We can also broaden and generalize our feature set to detect spam from any short URL services. In addition, we would like to develop a browser extension that can work in real time and classify any short link as malicious or benign. Appending a single plus sign to a Bitly URL gives the corresponding analytics, we can introduce a double plus sign here to get the credibility analytics for the same URL. Such a service can be very useful in short URL spam detection across multiple online media.

\end{document}